\documentclass[]{aa}
\usepackage{graphicx,natbib}

\def\aV{\mbox{$\rm A_V$}}

\def\jh{\mbox{$(J-H)$}}

\def\jk{\mbox{$(J-K_s)$}}
\def\mMJ{\mbox{$(m-M)_J$}}
\def\mMo{\mbox{$(m-M)_O$}}
\def\ebv{\mbox{$E(B-V)$}}

\def\ejh{\mbox{$E(J-H)$}}

\def\rc{\mbox{$R_{\rm c}$}}
\def\rch{\mbox{$R_{\rm c,H}$}}
\def\rl{\mbox{$R_{\rm RDP}$}}
\def\rx{\mbox{$R_{\rm ext}$}}

\def\rth{\mbox{$R_{\rm t,H}$}}

\def\ms{\mbox{$M_\odot$}}
\def\ds{\mbox{$d_\odot$}}
\def\rs{\mbox{$R_\odot$}}
\def\dgc{\mbox{$R_{\rm GC}$}}
\def\xgc{\mbox{$X_{\rm GC}$}}
\def\ygc{\mbox{$Y_{\rm GC}$}}
\def\zgc{\mbox{$Z_{\rm GC}$}}

\def\jj{\mbox{$J$}}
\def\hh{\mbox{$H$}}
\def\ks{\mbox{$K_s$}}

\def\ns{\mbox{$N_{1\sigma}$}}
\def\no{\mbox{$N_{\rm obs}$}}
\def\nc{\mbox{$N_{\rm cl}$}}
\def\sFS{\mbox{$\rm\sigma_{FS}$}}
\def\fsU{\mbox{$FS_{\rm unif}$}}

\def\tdis{\mbox{$t_{\rm dis}$}}

\begin{document}

\title{The old Galactic open clusters FSR\,1716 and Czernik\,23}

\author{C. Bonatto\inst{1} \and E. Bica\inst{1}}

\offprints{C. Bonatto}

\institute{Universidade Federal do Rio Grande do Sul, Departamento de Astronomia\\
CP\,15051, RS, Porto Alegre 91501-970, Brazil\\
\email{charles@if.ufrgs.br, bica@if.ufrgs.br}
\mail{charles@if.ufrgs.br} }

\date{Received --; accepted --}

\abstract
{Open clusters older than $\sim4$\,Gyr are rare in the Galaxy. Affected by a series of 
mass-decreasing processes, the stellar content of most open clusters dissolves into the field 
in a time-scale shorter than $\sim1$\,Gyr. In this sense, improving the statistics of old 
objects may provide constraints for a better understanding of the dynamical dissolution of open 
clusters.}
{Our main purpose is to investigate the nature of the Globular cluster candidate FSR\,1716, located
at $\ell=329.8^\circ$ and $b=-1.6^\circ$. We also derive parameters of the anti-centre open
cluster Czernik\,23 (FSR\,834). Both objects have been detected as stellar overdensities in the 
Froebrich, Scholz \& Raftery star cluster candidate catalogue.}
{The analyses are based on near-infrared colour-magnitude diagrams and stellar radial density
profiles. The intrinsic colour-magnitude diagram morphology is enhanced by a field-star
decontamination algorithm applied to the 2MASS \jj, \hh, and \ks\ photometry.}
{Isochrone fits indicate that FSR\,1716 is more probably an old ($\sim7$\,Gyr) and absorbed 
($\aV=6.3\pm0.2$) open cluster, located $\approx0.6$\,kpc inside the Solar circle in a contaminated 
central field. However, we cannot rule out the possibility of a low-mass, loose globular cluster.
Czernik\,23 is shown to be an almost absorption-free open cluster, $\sim5$\,Gyr old, located
about 2.5\,kpc towards the anti-centre. In both cases, Solar and sub-Solar ($[Fe/H]\sim-0.5$) metallicity
isochrones represent equally well the stellar sequences. Both star 
clusters have a low mass content ($\la200\,\ms$) presently stored in stars. Their relatively small 
core and cluster radii are comparable to those of other open clusters of similar age. These 
structural parameters are probably consequence of the several Gyrs of mass loss due to stellar 
evolution, tidal interactions with the disk (and bulge in the case of FSR\,1716), and possibly 
giant molecular clouds.}
{Czernik\,23, and especially FSR\,1716, are rare examples of extreme dynamical survivors. The
identification of both as such represents an increase of $\approx10\%$ to the known population
of open clusters older than $\sim4$\,Gyr in the Galaxy.}


\titlerunning{Very old OCs: FSR\,1716 and Cz\,23}

\maketitle

\section{Introduction}
\label{Intro}

The Galaxy is an aggressive environment to star clusters in general, the open clusters (OCs)
in particular. These stellar systems are continually harassed by a series of dynamical processes
such as mass loss associated to stellar evolution, mass segregation and evaporation, tidal
interactions with the Galactic disk and bulge, and collisions with giant molecular clouds.
Combined over time, such processes tend to accelerate the dynamical evolution, which produces
significant changes in the cluster structure and eroded mass functions. Eventually, most
OCs end up completely dissolved in the Galactic stellar field or as poorly-populated remnants
(\citealt{PB07} and references therein).

Theoretical (e.g. \citealt{Spitzer58}; \citealt{LG06}), N-body (e.g. \citealt{BM03}; \citealt{GoBa06};
\citealt{Khalisi07}), and observational (e.g. \citealt{vdB57}; \citealt{Oort58}; \citealt{vHoerner58};
\citealt{Piskunov07}) evidence indicate that the disruption-time scale near the Solar circle is shorter
than $\sim1$\,Gyr. Around this region, the disruption time-scale depends on mass as $\tdis\sim M^{0.62}$
(\citealt{LG06}), which for clusters with mass in the range $10^2 - 10^3\ms$ corresponds to
$\rm75\la\tdis(Myr)\la300$. In general, the effect of the relevant dynamical processes is stronger for
the OCs more centrally located and the low-mass ones (see \citealt{OldOCs} for a detailed discussion on
disruption effects and time-scales). Indeed, OCs older than $\sim1$\,Gyr are preferentially found near
the Solar circle and in the outer Galaxy (e.g. \citealt{Friel95}; \citealt{DiskProp}), where the frequency
of potentially damaging dynamical interactions with giant molecular clouds and the disk is lower (e.g.
\citealt{Salaris04}; \citealt{Upgren72}). Disruption efficiency increases critically towards the Galactic
centre, to the point that the inner ($\dgc\la150$\,pc) tidal fields can dissolve a massive star cluster
in $\sim50$\,Myr (\citealt{Portegies02}).

The above aspects considered, the natural expectation is that only a small fraction of the
OCs can reach old ages, and that the successful ones should be preferentially found at large
Galactocentric distances. In fact, of the $\approx1000$ OCs with known age listed in the
WEBDA\footnote{\em obswww.univie.ac.at/webda - \citet{Merm03}} database, 180 are older than 1\,Gyr,
and only 18 ($\approx2\,\%$) are older than 4\,Gyr (see also \citealt{OBB05a}; \citealt{OBB05b}).
Not surprisingly, most of the OCs older than 1\,Gyr so far identified are located outside the Solar 
circle (see, e.g., the spatial distribution of OCs of different ages in Fig.~1 of \citealt{OldOCs}).

\begin{figure*}
\begin{minipage}[b]{0.50\linewidth}
\includegraphics[width=\textwidth]{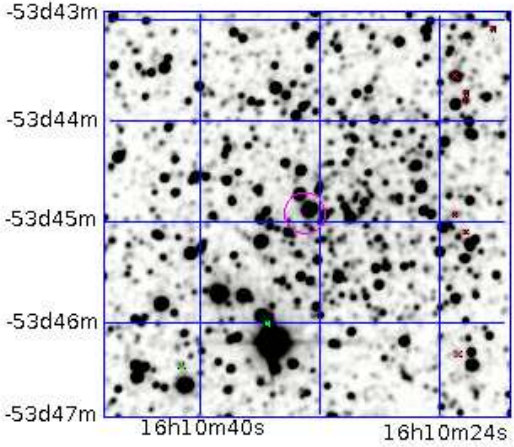}
\end{minipage}\hfill
\begin{minipage}[b]{0.50\linewidth}
\includegraphics[width=\textwidth]{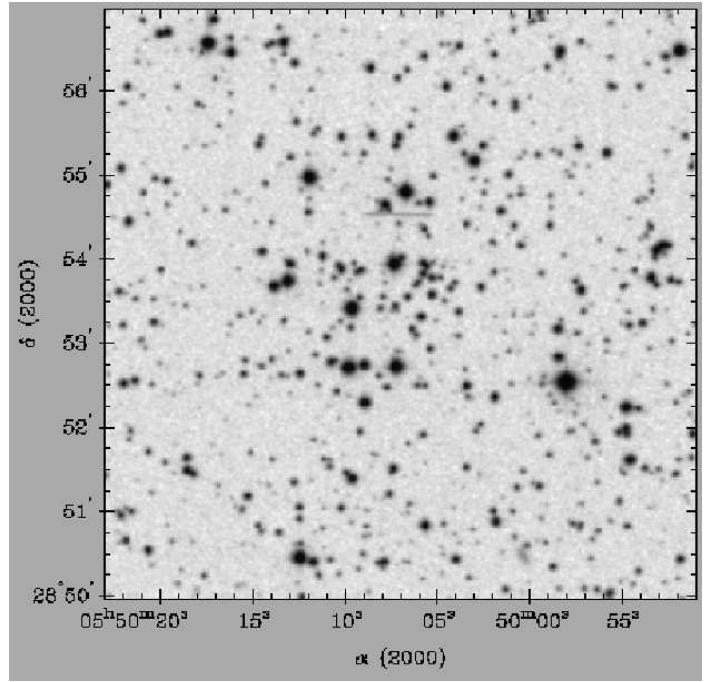}
\end{minipage}\hfill
\caption[]{Left panel: $4\arcmin\times4\arcmin$ 2MASS \ks\ image of FSR\,1716. Image provided by
the 2MASS Image Service. The small circle indicates the central coordinates (cols.~2 and 3 of
Table~\ref{tab1}). Right panel: $7\arcmin\times7\arcmin$ XDSS R image of Cz\,23. Figure orientation:
North to the top and East to the left.}
\label{fig1}
\end{figure*}

In this context, it is naturally expected that the discovery and derivation of astrophysical parameters
of old OCs will better define the OC parameter space. Thus, a better understanding on the dynamical
survival rate of star clusters in the tidal field of the Galaxy can be reached. Such parameters, in turn,
can be used in studies of star formation and evolution processes, dynamics of N-body systems, and the
geometry of the Galaxy, among others.

\begin{table*}
\caption[]{General data on the clusters}
\label{tab1}
\tiny
\renewcommand{\tabcolsep}{0.65mm}
\renewcommand{\arraystretch}{1.25}
\begin{tabular}{lcccccccccccccccc}
\hline\hline
&\multicolumn{7}{c}{FSR2007}&&\multicolumn{7}{c}{This paper}\\
\cline{2-8}\cline{10-16}
Cluster&$\alpha(2000)$&$\delta(2000)$&$\ell$&$b$&Q&\rch&\rth&&Age&\aV&\ds&\dgc&\xgc&\ygc&\zgc\\
&(hms)&($^\circ\,\arcmin\,\arcsec$)&($^\circ$)&($^\circ$)& &(\arcmin)&(\arcmin)
&&(Gyr)&(mag)&(kpc)&(kpc)& (kpc)&(kpc)& (kpc)\\
(1)&(2)&(3)&(4)&(5)&(6)&(7)&(8)&&(9)&(10)&(11)&(12)&(13)&(14)&(15)\\
\hline
FSR\,1716$^\dagger$&16:10:33&$-$53:44:12&329.79&$-1.59$&2&1.2&5.9&&$7.0\pm1.0$&$6.3\pm0.2$
      &$0.8\pm0.1$&$6.6\pm0.1$&$-6.6\pm0.1$&$-0.37\pm0.04$&$-0.02\pm0.01$\\
      
FSR\,1716$^\ddagger$&16:10:33&$-$53:44:12&329.79&$-1.59$&2&1.2&5.9&&$12.0\pm2.0$&$6.3\pm0.4$
      &$2.3\pm0.3$&$5.4\pm0.3$&$-5.3\pm0.3$&$-1.13\pm0.17$&$-0.06\pm0.01$\\
      
Cz\,23&05:50:07&$+$28:53:28&180.55&$+0.82$&1&0.8&4.2&&$5.0\pm1.0$&$0.0\pm0.1$
          &$2.5\pm0.1$&$9.7\pm0.2$  &$-9.7\pm0.1$&$-0.02\pm0.01$&$+0.04\pm0.01$\\
\hline
\end{tabular}
\begin{list}{Table Notes.}
\item Coordinates (cols.~2 to 5), quality flag (col.~6), and core and tidal radii (cols.~7 and 8)
measured in the \hh\ band are from \citet{FSRcat}; Col.~10: reddening 
towards the cluster's central region (Sect.~\ref{age}). Col.~11: distance from the Sun. Col.~12:
cluster Galactocentric distance for $\rs=7.2$\,kpc (\citealt{GCProp}). Cols.~13-15: coordinate
components projected onto the Galactic plane. Cz\,23 is the optical counterpart of FSR\,834.
Parameters of FSR\,1716 are derived for the OC ($\dagger$) or globular cluster ($\ddagger$) 
interpretation (Sect.~\ref{age}).
\end{list}
\end{table*}

In the present paper we study two stellar overdensities listed in the star cluster candidate catalogue 
of \citet{FSRcat}, which turn out to be very old star clusters. They are FSR\,1716 and FSR\,834. The
present work employs near-IR \jj, \hh, and \ks\ photometry obtained from the 2MASS\footnote{The Two
Micron All Sky Survey, All Sky data release (\citealt{2mass1997}), available at {\em
http://www.ipac.caltech.edu/2mass/releases/allsky/}} Point Source Catalogue (PSC). The spatial and
photometric uniformity of 2MASS, which allow extraction of wide surrounding fields that provide high
star-count statistics, are important to derive cluster parameters and probe the nature of stellar
overdensities (e.g. \citealt{ProbFSR}). For this purpose we have developed quantitative tools to
statistically disentangle cluster evolutionary sequences from field stars in colour-magnitude diagrams
(CMDs). Basically, we apply {\em (i)} field-star decontamination to quantify the statistical significance
of the CMD morphology, which is important to derive reddening, age, and distance from the Sun, and
{\em (ii)} colour-magnitude filters, which are essential for intrinsic stellar radial density profiles
(RDPs), as well as luminosity and mass functions (MFs). In particular, field-star decontamination
constrains more the age and distance, especially for low-latitude OCs (\citealt{DiskProp}).

This paper is organised as follows. Sect.~\ref{Target_OCs} contains basic properties and reviews
literature data (when available) on both star cluster candidates. In Sect.~\ref{2mass} we present
the 2MASS photometry and build the stellar surface-density distribution in the direction of both
objects. In Sect.~\ref{Decont_CMDs} we build CMDs, discuss the field-star decontamination algorithm,
and provide tests to the old age of both clusters. In Sect.~\ref{age} we derive cluster fundamental
parameters. Sect.~\ref{struc} describes cluster structure by means of stellar RDPs. In Sect.~\ref{MF}
mass functions are built and cluster masses are estimated. In Sect.~\ref{Discus} aspects related to
the structure and dynamical state of the present clusters are considered. Concluding remarks are given
in Sect.~\ref{Conclu}.

\section{FSR\,1716 and FSR\,834 as stellar overdensities}
\label{Target_OCs}

The catalogue built by \citet{FSRcat} includes 1021 star cluster candidates (hereafter FSR
objects) for Galactic latitudes $|b|<20^\circ$ and all longitudes. The targets were selected by 
an automated algorithm that basically identifies small-scale regions that present stellar overdensities,
applied to the 2MASS database. The overdensities are classified according to a quality flag, 
`0' and `1' for the most probable star clusters, while `5' and `6' may be related to field 
fluctuations. Based on a combination of the number of cluster stars (corrected to a common 
magnitude limit), the core radius and the central star density, \citet{FSRcat} could discriminate
known globular clusters (GCs) from OCs. With this criterion applied to the overdensity catalogue, 
they found 1012 OC and 9 GC candidates.

Several studies have explored the FSR catalogue with different approaches, with results that reflect the
importance of such catalogue. The recently discovered GCs FSR\,1735 (\citealt{FSR1735}) and FSR\,1767
(\citealt{FSR1767}), and the probable GCs FSR\,584 (\citealt{FSR584}) and FSR\,190 (\citealt{FSR190}),
are clear examples of the fundamental r\^ole played by the FSR catalogue to improve the statistics of
very old star clusters. Indeed, FSR\,1735 and FSR\,1767 are the most recent additions to the Galactic
GC population, a number that presently amounts to $\sim160$ members (e.g. \citealt{Pap11GCs}). On the
other hand, the decontamination algorithm (Sect.~\ref{Decont_CMDs}) applied to the 2MASS photometry of
some FSR GC candidates, has shown that FSR\,89 (\citealt{OldOCs}) and FSR\,1603 (\citealt{F1603}) are
open clusters of age $\sim1$\,Gyr, while FSR\,1755 appears to be an embedded OC (\citealt{F1603}).

During the course of a close investigation of the overdensities, we noted that the CMDs of FSR\,1716
and FSR\,834 show features typical of old stellar systems. Indeed, FSR\,1716 was classified as a GC
candidate by \citet{FSRcat}. FSR\,1716, at $\ell=329.79^\circ$ and $b=-1.59^\circ$ is projected not
far from the Galactic centre, which implies the presence of significant contamination by bulge stars.
FSR\,834, on the other hand, is a disk object projected towards the anti-centre ($\ell=180.55^\circ$
and $b=+0.82^\circ$). It has the poorly-studied OC Czernik\,23 (hereafter Cz\,23) as optical counterpart.
\citet{Czernik66} measured a diameter of 5\arcmin\ and estimated that Cz\,23 contains about 40 member 
stars, and \citet{Ruprecht66} which classified it as III\,1p.

Figure~\ref{fig1} (left panel) shows a $4\arcmin\times4\arcmin$ 2MASS \ks\ image of FSR\,1716, where
a significant concentration of stars is superimposed on a relatively crowded field, as expected from such 
a central direction. For Cz\,23 we show in the right panel a $7\arcmin\times7\arcmin$ XDSS\footnote{Extracted
from the Canadian Astronomy Data Centre (CADC), at \em http://cadcwww.dao.nrc.ca/} R band image. In this
case the cluster can be easily seen against a relatively clean field.

Table~\ref{tab1} provides fundamental data on both objects. Coordinates from \citet{FSRcat} are given in 
cols.~2 to 5; their quality flag in is col.~6, and the core and tidal radii measured in the \hh\ band
are given in cols.~7 and 8. The age, central reddening, distance from the Sun, Galactocentric distance,
and the components projected onto the Galactic plane derived in the present study (Sect.~\ref{age}) are 
given in Cols.~9 to 16.

\section{2MASS photometry}
\label{2mass}

2MASS photometry in the \jj, \hh, and \ks\ bands was extracted in circular fields of extraction radius
\rx\ centred on the coordinates of the objects (Table~\ref{tab1}) by means of VizieR\footnote{\em
http://vizier.u-strasbg.fr/viz-bin/VizieR?-source=II/246}. Previous studies with OCs in different 
environments (Sect.~\ref{Intro}) revealed that in the absence of a neighbouring populous cluster and
important differential absorption, wide extraction areas provide suitable statistics for a
consistent characterisation of the field stars in terms of colour and magnitude. Based on this premise
we adopted an extraction radius of $\rx=30\arcmin$, which is beyond the cluster radius
(Sect.~\ref{struc} and col.~9 of Table~\ref{tab3}) of the present objects. For decontamination purposes, 
comparison fields were extracted within wide rings located beyond the cluster radii. As a photometric
quality constraint, the 2MASS extractions were restricted to stars {\em (i)} brighter than 
those of the 99.9\% Point Source Catalogue completeness 
limit\footnote{\em http://www.ipac.caltech.edu/2mass/releases/allsky/doc/ } in the cluster
direction, and {\em (ii)} with errors in \jj, \hh, and \ks\ smaller than 0.3\,mag. The 99.9\% completeness
limits refer to field stars, and depend on Galactic coordinates. In the present cases, the fraction
of stars with \jj, \hh, and \ks\ uncertainties smaller than 0.06\,mag is $\approx80\%$. A typical distribution
of uncertainties as a function of magnitude, for objects projected towards the central parts of the Galaxy, 
can be found in \citet{BB07}. Reddening transformations use the relations $A_J/A_V=0.276$, $A_H/A_V=0.176$,
$A_{K_S}/A_V=0.118$, and $A_J=2.76\times\ejh$ (\citealt{DSB2002}), for a constant total-to-selective 
absorption ratio $R_V=3.1$. These ratios were derived from the extinction curve of \citet{Cardelli89}.

\begin{figure}
\resizebox{\hsize}{!}{\includegraphics{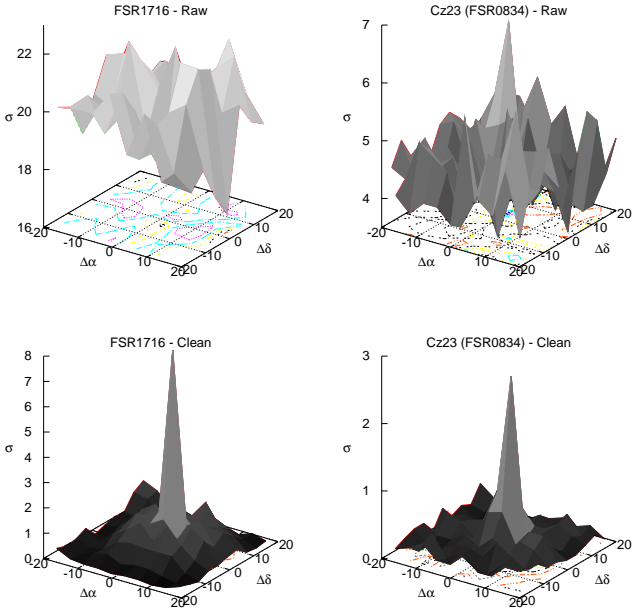}}
\caption[]{Stellar surface-density $\sigma(\rm stars\ arcmin^{-2})$ of FSR\,1716 (left panels)
and Cz\,23 (right). The curves were computed for a mesh size of $4\arcmin\times 4\arcmin$, centred
on the coordinates in Table~\ref{tab1}. The observed (raw) and field-star decontaminated photometry
are shown in the top and bottom panels, respectively.}
\label{fig2}
\end{figure}

\begin{figure}
\resizebox{\hsize}{!}{\includegraphics{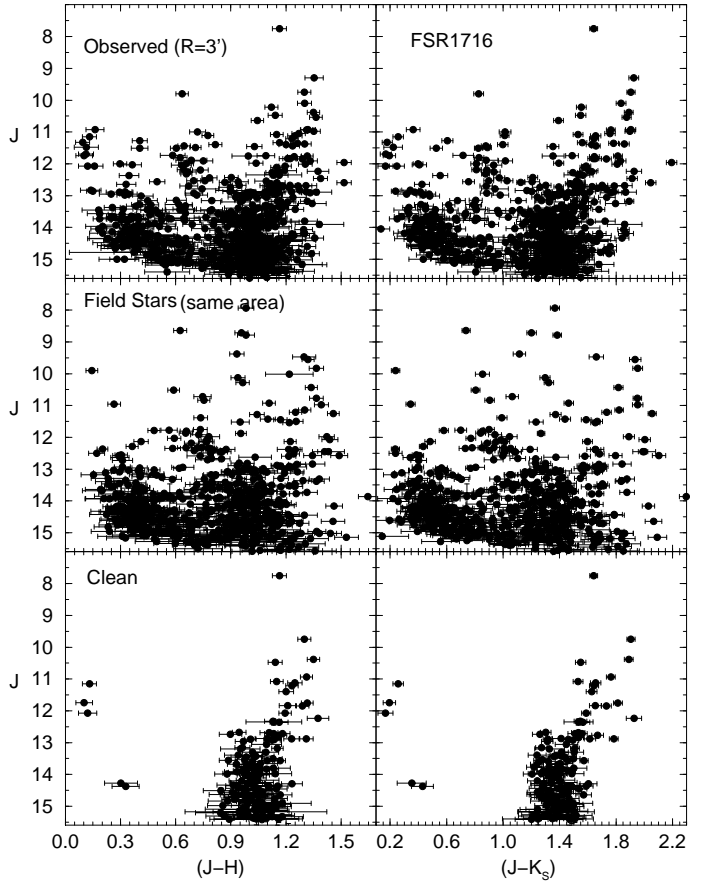}}
\caption[]{2MASS CMDs extracted from the $R<3\arcmin$ region of FSR\,1716. Top panels:
observed photometry with the colours $\jj\times\jh$ (left) and $\jj\times\jk$ (right). Middle:
equal-area ($19.77\arcmin<R<20\arcmin$) extraction of the comparison field, where the important 
disk and bulge contamination can be seen. Bottom panels: decontaminated CMDs that reveal 
a highly reddened, relatively populous MS and a giant branch, typical of old clusters.}
\label{fig3}
\end{figure}

\begin{figure}
\resizebox{\hsize}{!}{\includegraphics{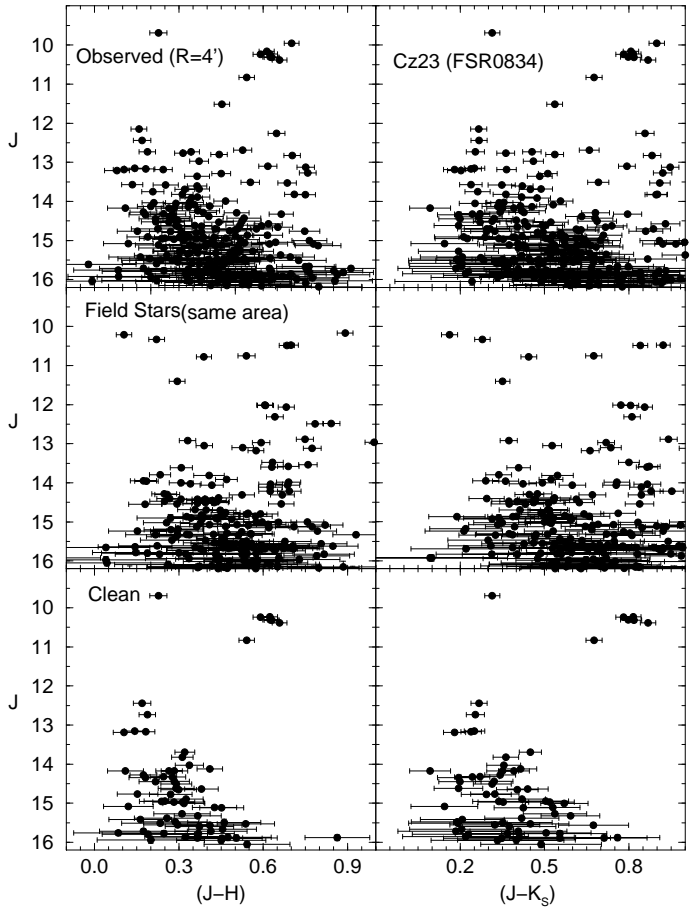}}
\caption[]{Same as Fig.~\ref{fig3} for the region $R<4\arcmin$ of Cz\,23. Contamination, in this 
case, is less important. The equal-area comparison field extraction was taken from the region
$19.6\arcmin<R<20\arcmin$. A giant clump shows up especially in the decontaminated CMDs, denoting 
advanced age.}
\label{fig4}
\end{figure}

\begin{figure}
\resizebox{\hsize}{!}{\includegraphics{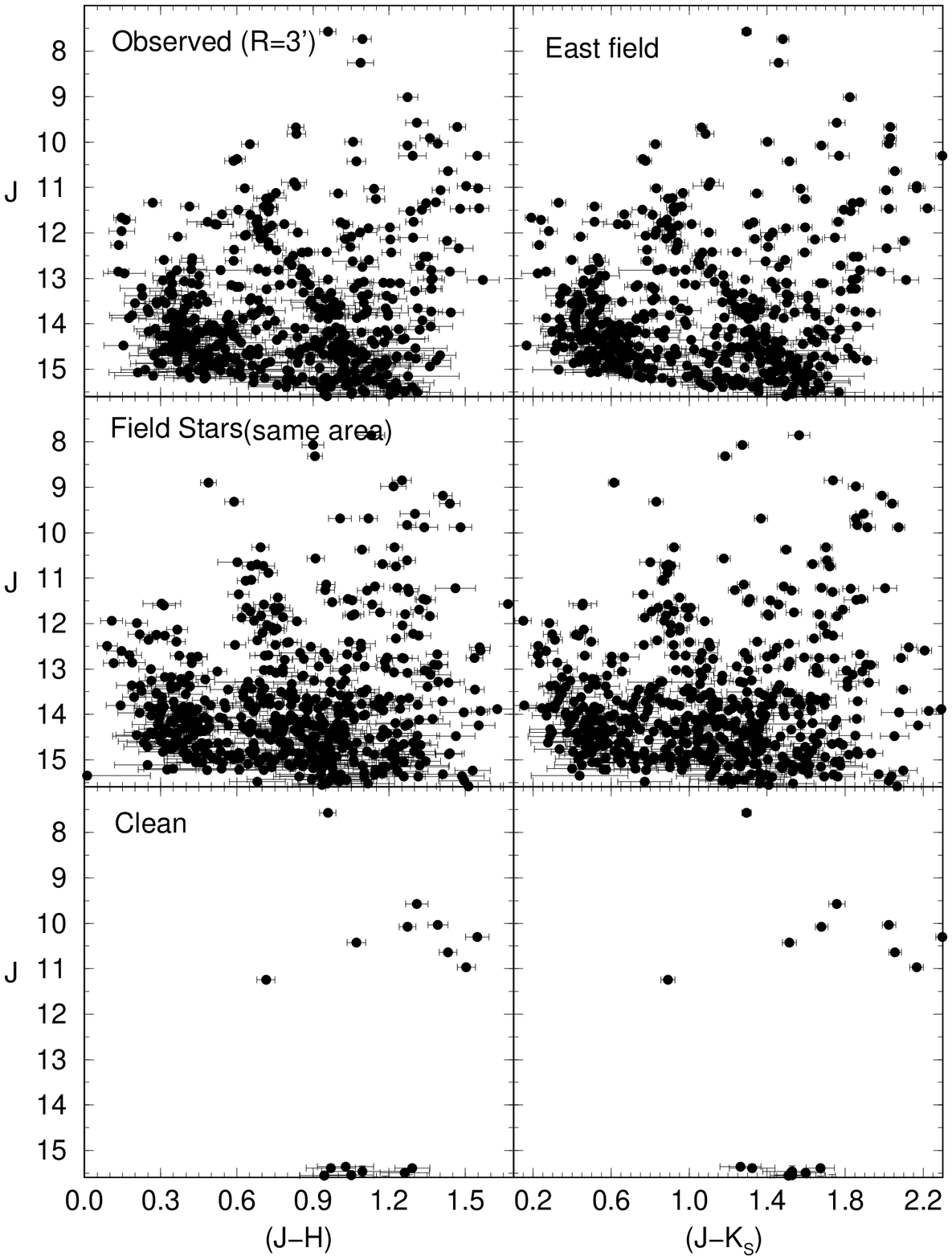}}
\caption[]{Same as Fig.~\ref{fig3} for the $R<3\arcmin$ region of a 'test-field' located at
$1^\circ$ to the East of FSR\,1716. The decontaminated CMDs (bottom panels) contain essentially
statistical noise.}
\label{fig5}
\end{figure}

As a first step to understand the nature of both overdensities, we show in Fig.~\ref{fig2} the
spatial distribution of the stellar surface-density as measured with 2MASS photometry. In both cases
we compute the surface density ($\sigma$, in units of $\rm stars\,arcmin^{-2}$) in a rectangular
mesh with cells of dimensions $4\arcmin\times4\arcmin$. The mesh reaches total offsets of
$|\Delta\alpha|=|\Delta\delta|\approx20\arcmin$ with respect to the centre, in right ascension and 
declination. Since the cluster radius of both objects is $\la6\arcmin$ (Table~\ref{tab3}), most of 
the cluster structure should be contained in the central cell. Because of the important contamination 
by field stars, the surface-density built with the observed (raw) photometry of the centrally projected
OC FSR\,1716 is very irregular (top-left panel), although an excess can be seen in the central cell, 
which corresponds to the overdensity detected by \citet{FSRcat}. Cz\,23, on the other hand, clearly 
detaches in the central cell (top-right panel) against a less-irregular field. The bottom panels show 
the surface densities built with field-star decontaminated photometry (Sect.~\ref{Decont_CMDs}). 

\section{Field-star decontamination}
\label{Decont_CMDs}

Especially in the case of FSR\,1716, the stellar surface-density in the direction of both objects
clearly shows that field-star contamination should be taken into account. This assertion is further
supported by the $\jj\times\jh$ and $\jj\times\jk$ CMDs extracted from the $\rm R<3\arcmin$ region
of FSR\,1716  (Fig.~\ref{fig3}), and the $\rm R<4\arcmin$ region of Cz\,23 (Fig.~\ref{fig4}). Features 
present in the central CMDs and in the respective comparison field (top and middle panels), show that 
field stars contribute in varying proportions to the CMDs, especially for the bulge-projected FSR\,1716. 
Nevertheless, when contrasted with the equal-area comparison field extractions, sequences typical of old
stellar systems are suggested in both cases, a conspicuous giant branch in FSR\,1716 and a giant clump
in Cz\,23.

Field-star decontamination is a very important, yet difficult, step in the identification and
characterisation of star clusters. Different approaches have been used to this purpose, among them, 
those of \citet{Mercer05} and \citet{Carraro06}. The first is based essentially
on spatial variations of the star-count density, but does not take into account CMD properties. In the
latter, stars in a CMD extracted from an assumed cluster region are subtracted according to colour and
magnitude similarity with the stars of an equal-area comparison field CMD.

In the present case, we apply the statistical algorithm described in \citet{BB07} to quantify the
field-star contamination in the CMDs. The algorithm makes use of both approaches above, in the sense that
relative star-count density together with colour/magnitude similarity between cluster and comparison field
extractions are taken into account simultaneously. It measures the relative number densities of probable
field and cluster stars in cubic CMD cells whose axes correspond to the \jj\ magnitude and the \jh\ and 
\jk\ colours. These are the 2MASS colours that provide the maximum variance among CMD sequences for OCs 
of different ages (e.g. \citealt{TheoretIsoc}). The algorithm: {\em (i)} divides the full range of 
magnitude and colours covered by the CMD into a 3D grid, {\em (ii)} calculates the expected number 
density of field stars in each cell based on the number of comparison field stars with similar magnitude 
and colours as those in the cell, and {\em (iii)} subtracts the expected number of field stars from each 
cell. The algorithm is responsive to local variations of field-star contamination (\citealt{BB07}). 
Typical cell dimensions are $\Delta\jj=1.0$, and $\Delta\jh=\Delta\jk=0.25$, which are wide enough to
allow for sufficient star-count statistics in individual cells and narrow enough to preserve the
morphology of the CMD evolutionary sequences. The comparison fields are located within
$\rm R=15\arcmin - 30\arcmin$ (FSR\,1716), and $\rm R=10\arcmin - 30\arcmin$ (Cz\,23). In
both cases, the inner boundary of the comparison field lies beyond the probable tidal radius
(Sect.~\ref{struc}), which minimises the probability of oversubtraction of member stars. We emphasise that 
the equal-area field extractions shown in the middle panels of Figs.~\ref{fig3} and \ref{fig4} serve only for
comparisons among the panels. Actually, the decontamination process is carried out with the wide surrounding
area as described above. Further details on the algorithm, including discussions on subtraction efficiency
and limitations, are given in \citet{BB07}.

As extensively discussed in \citet{BB07}, differential reddening between cluster and field stars
is really critical for the decontamination algorithm. Large gradients would require wide cell sizes
or, in extreme cases, would preclude application of the algorithm altogether. Basically, it would be
required, e.g. $|\Delta\jh|\ga\rm cell~size$ (0.25, in the present work) between cluster and comparison
field for the differential reddening to affect the subtraction in a given cell. However, in both cases
the CMDs extracted from the cluster region and comparison field (Figs.~\ref{fig3} and \ref{fig4})
indicate that the differential reddening, if present, is not important.

The decontaminated stellar surface-density distributions (bottom panels of Fig.~\ref{fig2}) are 
an indicator of the algorithm efficiency. In both cases the central cells, which 
correspond to the location of the overdensities, present conspicuous surface densities. Besides, the 
residual surface-density around the centre was reduced to a minimum level. We note that the decontamination
is essentially based on the colour-magnitude distributions of the stars in different spatial regions.
The fact that the decontaminated surface density ends up with a conspicuous excess only at the assumed 
cluster region (Fig.~\ref{fig2}) implies significant differences among the spatial regions, both in the
colour-magnitude distributions and in the number of stars within a given colour-magnitude bin. This is 
expected of star clusters, which can be basically characterised by a single-stellar population, projected 
against a Galactic stellar field.

The decontaminated CMDs are shown in the bottom panels of Figs.~\ref{fig3} and \ref{fig4}. As expected, 
essentially all of the bulge and disk contamination in FSR\,1716 is removed, leaving stellar sequences 
typical of a reddened old OC, with a rather populous main sequence (MS), well-defined main sequence 
turnoff (MSTO), and 
the giant branch. Alternatively, the decontaminated morphology might resemble that of a poorly-populated
GC. We explore this possibility in Sect.~\ref{NatureFSR1716}. Of the 588 stars present in the $R<3\arcmin$ CMD, only 
128 remain. A similar conclusion 
applies to Cz\,23, in which the disk contamination has been subtracted, revealing a poorly-populated, old 
OC; 66 of the 228 stars remain in the CMD after decontamination. 

The approximately central direction of FSR\,1716 and the poorly-populated nature of Cz\,23 require 
additional statistical analysis. For this purpose, we present in Table~\ref{tab2} the full statistics 
of the decontaminated sequences and field stars, discriminated by magnitude bins. Statistically relevant 
parameters that characterise the nature of a star cluster are: {\em (i)} \ns\ which,
for a given magnitude bin, corresponds to the ratio of the decontaminated number of stars to the $\rm1\sigma$
Poisson fluctuation of the number of observed stars, {\em (ii)} \sFS, which is related to the probability
that the decontaminated stars result from the normal star count fluctuation in the comparison field
and, {\em (iii)} \fsU, which measures the star-count uniformity of the comparison field. Properties of \ns,
\sFS, and \fsU, measured in OCs and field fluctuations are discussed in \citet{ProbFSR}. Table~\ref{tab2}
also provides integrated values of the above parameters, which correspond to the full magnitude range spanned
by the CMD of each OC. The spatial regions considered here are those sampled by the CMDs shown in the top panels 
of Figs.~\ref{fig3} and \ref{fig4}.

Star cluster CMDs should have integrated \ns\ values significantly higher than 1 (\citealt{ProbFSR}),
a condition that is met by FSR\,1716 ($\ns=5.0$) and Cz\,23 ($\ns=4.4$). Besides, the number of decontaminated
stars in each magnitude bin of FSR\,1716 is higher (at the $3\sigma$ level or higher) than what could be
expected from field-star fluctuations. As a further test of the statistical significance of the above results
we investigate star count properties of the field stars. First, the comparison field is divided into 8
sectors around the cluster centre. Next, we compute the parameter
\sFS, which is the $\rm 1\,\sigma$ Poisson fluctuation around the mean of the star counts measured in the 8
sectors of the comparison field (corrected for the different areas of the sectors and cluster extraction). In
a spatially uniform comparison field, \sFS\ is expected to be low. In this context, star clusters should have
the probable number of member stars (\nc) higher than $\sim3\,\sFS$, to minimise the probability that \nc\
arises from fluctuations of a non-uniform comparison field. This condition is fully satisfied, in some cases
reaching the level $\nc\sim5\,\sFS$. The ratio decreases somewhat for Cz\,23, probably because it has a low
number of member stars. Finally, we also provide in Table~\ref{tab2} the parameter \fsU.
For a given magnitude bin we first compute the average number of stars over all sectors $\langle N\rangle$
and the corresponding $\rm 1\sigma$ fluctuation $\sigma_{\langle N\rangle}$; thus, \fsU\ is defined as
$\rm\fsU=\sigma_{\langle N\rangle}/\langle N\rangle$. Non uniformities such as heavy differential reddening
should result in high values of \fsU.

Since we usually work with comparison fields wider than the possible-cluster extractions, the correction
for the different spatial areas between field and cluster is expected to produce a fractional number
of probable field stars ($n_{fs\_exp}^{cell}$) in some cells. Before the cell-by-cell subtraction, the
fractional numbers are rounded off to the nearest integer, but limited to the number of observed stars
in each cell $n_{fs\_sub}^{cell}=NI(n_{fs\_exp}^{cell})\leq n_{obs}^{cell}$, where NI represents the nearest
integer. The global effect is quantified by means of the difference between the expected number
of field stars in each cell ($n_{fs\_exp}^{cell}$) and the actual number of subtracted stars ($n_{fs\_sub}^{cell}$).
Summed over all cells, this quantity provides an estimate of the total subtraction efficiency of the
process, \[ f_{sub}=100\times\sum_{cell}n_{fs\_sub}^{cell}/\sum_{cell}n_{fs\_exp}^{cell}~~~(\%).\] Ideally, the best
results would be obtained for an efficiency $f_{sub}\approx100\%$. With the assumed grid settings for the
decontamination of FSR\,1716 and Cz\,23, the subtraction efficiencies turned out to be higher than 93\%.

\begin{table*}
\caption[]{Statistics of the field-star decontamination discriminated by magnitude bins}
\label{tab2}
\renewcommand{\tabcolsep}{2.85mm}
\renewcommand{\arraystretch}{1.2}
\begin{tabular}{ccccccccccccccc}
\hline\hline
$\Delta\jj$&&\multicolumn{5}{c}{FSR\,1716 ($R<3\arcmin$)}&&&\multicolumn{5}{c}{Cz\,23 ($R<4\arcmin$)}\\
\cline{3-7}\cline{10-14}
 &&\no&\nc&\ns&\sFS&\fsU& & &\no&\nc&\ns&\sFS&\fsU\\
(mag)&&(stars)&(stars)&&(stars)&&&&(stars)&(stars)&&(stars)\\
\cline{1-14}
  7--8&&$1\pm1.0$  &1 &1.0&0.20&0.36& & &---&---&---&---&---    \\
  8--9&&$1\pm1.0$  &1 &1.0&0.38&0.23& & &---&---&---&---&---    \\
 9--10&&$3\pm1.7$  &1 &0.6&0.37&0.08& & &$2\pm1.4$   &1 &0.7&0.47&0.63    \\
10--11&&$11\pm3.3$ &3 &0.9&0.81&0.07& & &$7\pm2.6$   &6 &2.3&0.91&0.38 \\
11--12&&$37\pm6.1$ &9 &1.5&1.90&0.07& & &$1\pm1.0$   &0 &0.0&1.08&0.22 \\
12--13&&$82\pm9.1$ &20&2.2&3.23&0.05&&  &$10\pm3.2$  &2 &0.6&1.50&0.15  \\
13--14&&$145\pm12.0$&36&3.0&6.33&0.05&& &$24\pm4.9$  &5 &1.0&2.33&0.11  \\
14--15&&$248\pm15.7$&57&3.6&8.45&0.04&& &$66\pm8.1$  &22&2.7&3.77&0.08  \\
15--16&&---&---&---&---&---&          & &$118\pm10.9$&30&2.8&5.61&0.07  \\
\cline{3-7}\cline{10-14}
All &&$528\pm23.0$&128&5.0&19.2&0.04&& &$228\pm15.1$&66&4.4&15.6&0.07 \\
\hline
\end{tabular}
\begin{list}{Table Notes.}
\item The table provides, for each magnitude bin ($\Delta\jj$), the number of observed stars 
(\no) within the spatial region sampled in the CMDs shown in the top panels of Figs.~\ref{fig3} 
and \ref{fig4}, the respective number of probable member stars (\nc) computed by the 
decontamination algorithm, the \ns\ parameter, the $\rm 1\,\sigma$ Poisson fluctuation (\sFS) around 
the mean, with respect to the star counts measured in the 8 sectors of the comparison field, and the 
field-star uniformity parameter. The statistical significance of \nc\ is reflected in its ratio with 
the $1\sigma$ Poisson fluctuation of \no\ (\ns) and with \sFS. The bottom line corresponds to the  
full magnitude range.
\end{list}
\end{table*}

\begin{figure}
\resizebox{\hsize}{!}{\includegraphics{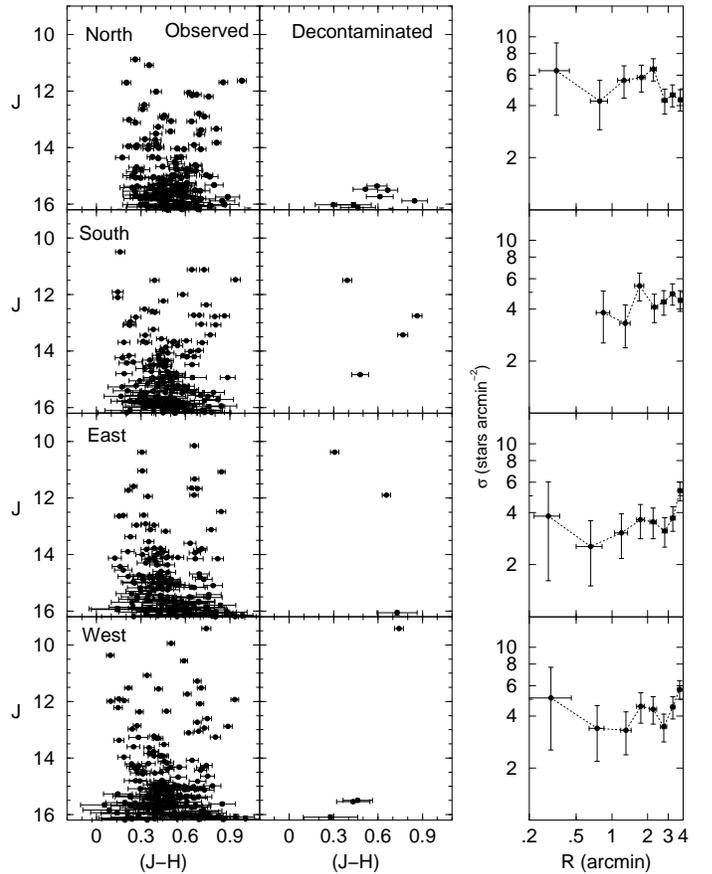}}
\caption[]{Left panels: Observed and decontaminated CMDs extracted from $R=4\arcmin$ comparison fields,
taken at 20\arcmin\ to each side of Cz\,23. The corresponding RDPs (right panels) exhibit essentially
field-fluctuation features.}
\label{fig6}
\end{figure}

Finally, it's worth noting that the qualitative and quantitative expectations of the
decontamination algorithm have been met by the output. On the one hand, the decontaminated
photometry reveals a very-high excess in the surface-density distributions with respect to the
surroundings, in both cases (Fig.~\ref{fig2}). In addition, CMDs extracted from the spatial regions
where the excesses occur present statistical significant (Table~\ref{tab2}) cluster-like features
(Figs.~\ref{fig3} and \ref{fig4}).

\subsection{Additional test for FSR\,1716: field at $1^\circ$ to the East}
\label{EastField}

Evidence drawn from the previous sections indicates that FSR\,1716 is an old star cluster. However, 
since it lies at a low latitude and is projected not far from the dense bulge stellar field, we provide 
an additional test to further probe its nature. A `test-field' of 20\arcmin\ of radius was extracted at
$1^\circ$ to the East of FSR\,1716, with Galactic coordinates $\ell=330.79^\circ$ and $b=-1.59^\circ$,
thus projected somewhat closer to the centre than FSR\,1716. The 'test-field' photometry was analysed
in the same way as that of FSR\,1716. Similarly to FSR\,1716, we considered a central region ($R<3\arcmin$) 
of the `test-field', extracted the equal-area comparison field near the border ($19.77\arcmin - 20\arcmin$), 
and applied the decontamination algorithm. The results are shown in Fig.~\ref{fig5}. Contrarily to the 
old-cluster decontaminated CMD morphology of FSR\,1716 (Fig.~\ref{fig3}), the decontaminated CMD of the 
`test-field' (bottom panels of Fig.~\ref{fig5}) essentially contains Poisson noise produced by the mutual 
subtraction of star samples that share similar distributions of magnitude and colours. The same conclusion 
can be drawn from the featureless `test-field' RDP (Sect.~\ref{struc}; Fig.~\ref{fig10}).

As discussed in \citet{ProbFSR}, Poisson fluctuations of the dense stellar field projected towards the
bulge do not produce cluster-like CMDs and RDPs simultaneously. In particular, when field-fluctuation
CMDs are field-star decontaminated, what results is essentially represented by the above FSR\,1716
`test-field' experiment (Fig.~\ref{fig5}). In this sense, the striking differences exhibited by the CMD 
and RDP of FSR\,1716, as compared to those of the `test-field', can be taken as a robust evidence of the 
old-cluster nature of FSR\,1716.

\subsection{Additional comparison fields for Cz\,23}
\label{CompFields}

The CMD of Cz\,23, especially the decontaminated one (Fig.~\ref{fig4}), suggests a poorly-populated 
old OC. Thus, we test whether the decontamination procedure, applied to random fields around Cz\,23,
may produce similar cluster-like sequences and stellar radial profiles. We consider $R=4\arcmin$ 
comparison fields, taken at 20\arcmin\ to each side of Cz\,23. These fields have the same projected 
area as that of Cz\,23 shown in Fig.~\ref{fig4}, and the centre to centre offsets correspond to 
$\approx4$ times the cluster radius of Cz\,23 (Table~\ref{tab3}). Field-star decontamination was applied 
to these CMDs using the same offset field as that employed for Cz\,23 (Sect.~\ref{Decont_CMDs}),
but restricted to $R=25\arcmin - 30\arcmin$ to avoid self subtraction.

The observed and decontaminated comparison field CMDs are shown in Fig.~\ref{fig6}. In stark contrast 
with the decontaminated CMD of Cz\,23 (Fig.~\ref{fig4}), what remains in the $R=4\arcmin$ comparison 
field CMDs is a randomly scattered small number of stars, similar to the CMD of the FSR\,1716 `test-field' 
(Sect.~\ref{EastField}). To complete this analysis we also include in Fig.~\ref{fig6} the RDPs of the 
comparison fields. Both CMDs and RDPs of the comparison fields are typical of Poisson fluctuations of 
stellar fields projected towards the Galactic anti-centre (\citealt{AntiCent}).

As expected of field fluctuations, the CMDs and RDPs of the comparison fields (Fig.~\ref{fig6}) are
featureless, quite contrasting with the cluster CMD (Fig.~\ref{fig4}) and RDP (Fig.~\ref{fig11})
of Cz\,23.

\begin{figure}
\resizebox{\hsize}{!}{\includegraphics{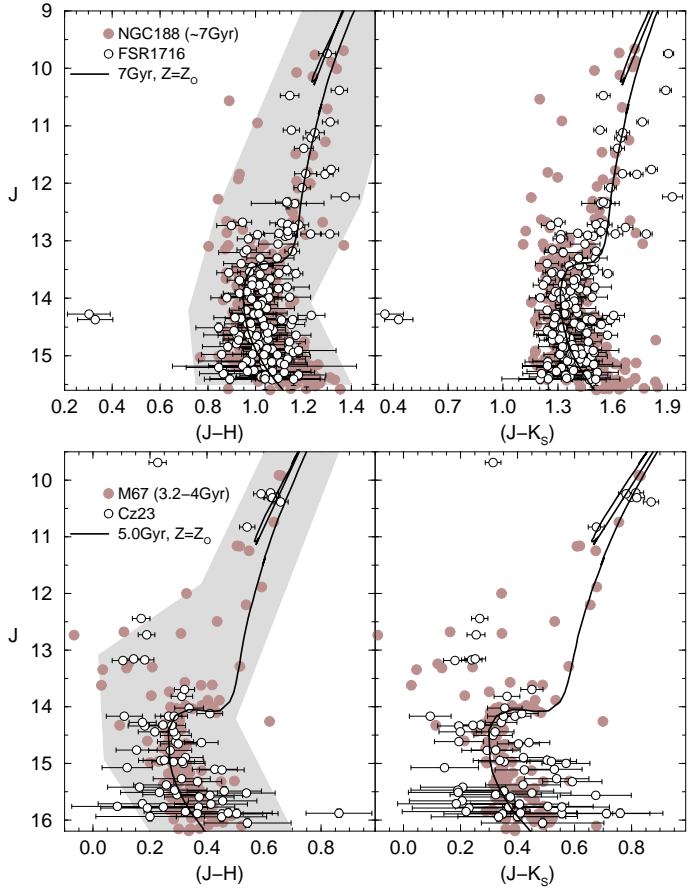}}
\caption[]{Top panels: the field-decontaminated ($R<3\arcmin$) CMD morphology of FSR\,1716 (empty 
circles) is compared to that of the $\sim7$\,Gyr OC NGC\,188 (gray circles). $\jj\times\jh$ (left 
panels) and $\jj\times\jk$ (right) CMDs are shown. Bottom: same as above for the $R<4\arcmin$ CMD 
of Cz\,23 (empty circles) and M\,67 (gray circles). The {\em adopted fit} (Sect.~\ref{ageFSR}) for 
FSR\,1716 corresponds to the 7\,Gyr, Solar-metallicity Padova isochrone, while for Cz\,23, we adopted 
the 5\,Gyr, Solar-metallicity isochrone (Sect.~\ref{ageCz}). The shaded polygon overplotted in the 
left panels shows the colour-magnitude filter. }
\label{fig7}
\end{figure}

\begin{figure}
\resizebox{\hsize}{!}{\includegraphics{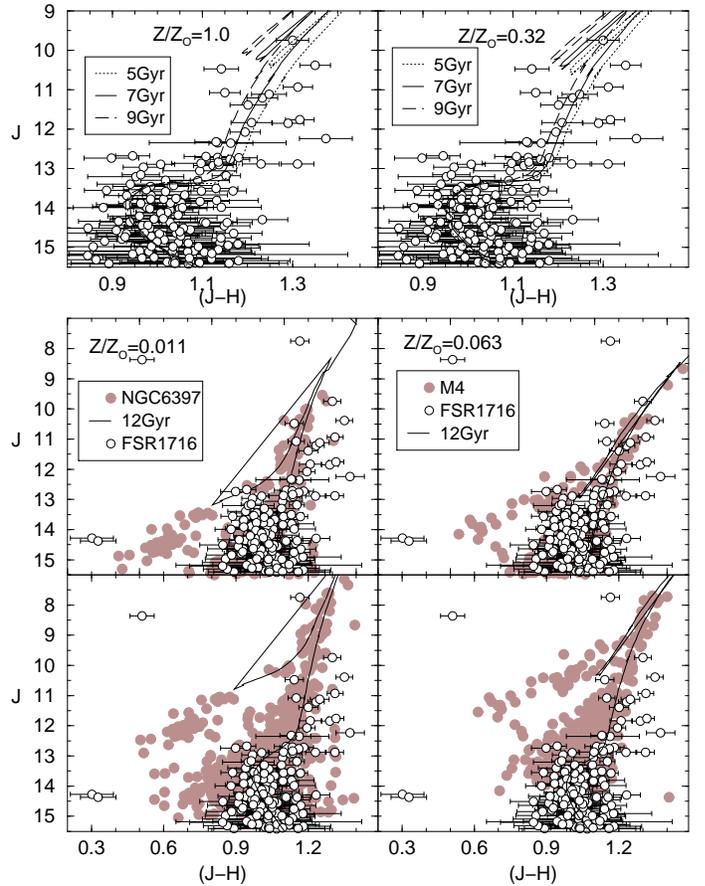}}
\caption[]{Age and metallicity interpretation of FSR\,1716. Top-left panel: isochrones of different 
ages and fixed Solar metallicity. Top-right: same ages but with the sub-Solar metallicity
$[Fe/H]=-0.5$ ($Z=0.006$).  Fits have in common a similar description of the MS. Middle panels:
template fit assuming a GC-like RGB and sub-giant branch morphology for FSR\,1716, with the GC 
NGC\,6397 ($[Fe/H]=-1.95$, $Z=0.0002$ - left panel), and M\,4 ($[Fe/H]=-1.21$, $Z=0.0012$ - right). 
Padova isochrones of 12\,Gyr and same metallicity as the GCs are shown. Bottom panels: same as above
but assuming a brighter MSTO for FSR\,1716. The colour range in each panel is optimised to shown 
differences among the fits (top) and enhance the relevant features (bottom).}
\label{fig8}
\end{figure}

\begin{figure}
\resizebox{\hsize}{!}{\includegraphics{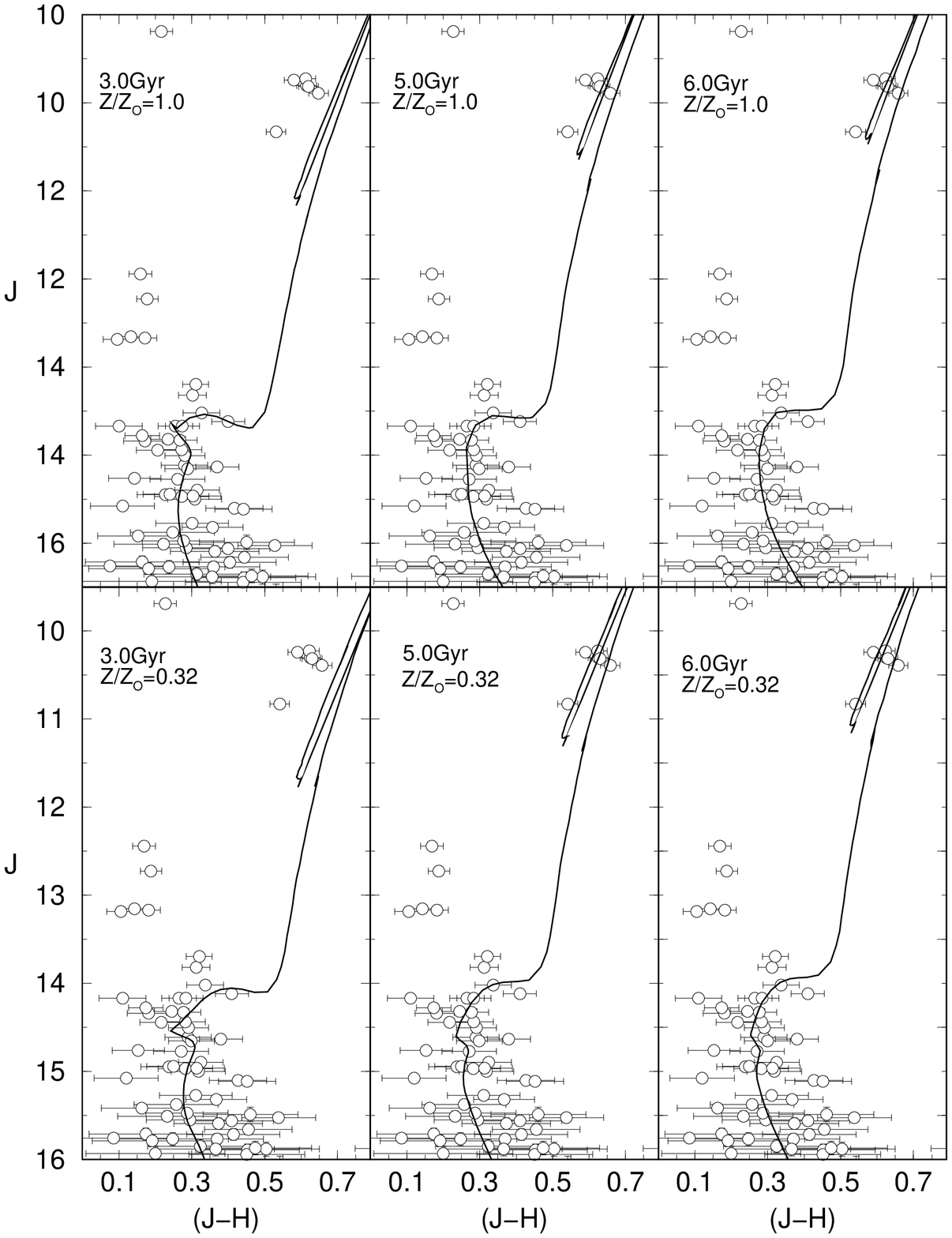}}
\caption[]{Age and metallicity interpretation of Cz\,23. Top panels: isochrones of different 
ages and fixed Solar metallicity. Bottom: same ages but with the sub-Solar metallicity 
$[Fe/H]=-0.5$ ($Z=0.006$). Fits are required to provide a similar description of the MS.}
\label{fig9}
\end{figure}

\section{Age, reddening, and distance}
\label{age}

Having established that FSR\,1716 and Cz\,23 present old-age features, we proceed to derive 
their fundamental parameters. We work with Padova isochrones (\citealt{Girardi2002}) of Solar 
and sub-Solar metallicity, computed with the 2MASS \jj, \hh, and \ks\ filters\footnote{\em http://stev.oapd.inaf.it/cgi-bin/cmd}.
These isochrones are very similar to the Johnson-Kron-Cousins ones (e.g. \citealt{BesBret88}), with
differences of at most 0.01 in \jh\ (\citealt{TheoretIsoc}). We adopt $\rs=7.2\pm0.3$\,kpc (\citealt{GCProp})
as the Sun's distance to the Galactic centre to compute the OC's Galactocentric distances. The value
$\rs=7.2\pm0.3$\,kpc was derived by means of the GC spatial distribution. Other recent studies
gave similar results, e.g. $\rs=7.2\pm0.9$\,kpc (\citealt{Eisen03}), $\rs=7.62\pm0.32$\,kpc
(\citealt{Eisen05}) and $\rs=7.52\pm0.10$\,kpc (\citealt{Nishiyama06}), with different approaches.
We make fits {\em by eye}, taking into account that, because of the presence of binaries, the 
isochrone should be shifted to the left of the MS fiducial line (e.g. \citealt{N188}, and references 
therein).

In principle, the decontaminated CMD morphologies (bottom panels of Figs.~\ref{fig3} and \ref{fig4})
- as well as the additional tests discussed in Sect.~\ref{EastField} - should provide enough
constraints to derive reliable cluster fundamental parameters. However, in view of the importance of
identifying new old star clusters, we explore a range of possibilities in terms of age and metallicity.
As discussed by, e.g. \citet{Friel95}, OC metallicities in general range from Solar ($[Fe/H]=0$, 
or $Z=0.019$) to sub-Solar ($[Fe/H]=-0.5$, $Z=0.006$) values. 

\subsection{CMD morphology of FSR\,1716}
\label{NatureFSR1716}

First, we compare in Fig.~\ref{fig7} (top panels) the decontaminated CMD of FSR\,1716 to that of 
the $\sim7$\,Gyr OC NGC\,188 (\citealt{N188}, and references therein). CMD morphologies in 
both colours show a good agreement along the MS, MSTO, giant branch, and the MS width, which  
suggests that both clusters are of similar age. In particular, both CMDs present a similar  
smoothly-curved MSTO and a scarcely populated red clump.

Different ages and metallicities are tested in Fig.~\ref{fig8}. Solar-metallicity solutions 
with isochrones of ages 5, 7, and 9\,Gyr are shown in the top-left panel. Same ages but with 
the low OC metallicity range ($[Fe/H]\sim-0.5$, which corresponds to $Z\approx0.006$), are 
considered in the top-right panels. As fit constraint, the isochrones have in common a similar 
description of the MS. In both panels, the youngest isochrone produces a relatively poorer fit, 
especially to the giant branch. On the other hand, the older isochrones - of both metallicity 
ranges - produce qualitatively similar fits. This not only confirms the old age of FSR\,1716, 
but, together with a few blue stars at $\jj\approx14.3$ and $\jh\approx0.3$ (Fig.~\ref{fig7}),
which might suggest a blue horizontal branch (HB), raise the possibility of a poorly-populated, 
metal-poor GC. 

To test the GC hypothesis we compare the decontaminated CMD of FSR\,1716 to those of the blue HB 
GCs NGC\,6397 and M\,4 (NGC\,6121). Both GCs are among the nearest ones ($\ds\approx2.3$\,kpc), 
which provides MS depth for 2MASS photometry, and are metal-poor, with $[Fe/H]\sim-1.95$ (NGC\,6397) 
and $[Fe/H]\sim-1.2$ (M\,4). We also use Padova isochrones of 12\,Gyr to characterise the GC ages 
(e.g. \citealt{OrtoNat}). The CMDs of NGC\,6397 (extracted within $\rx<3\arcmin$) and M\,4 
($\rx<2\arcmin$) were produced in the same way as those of FSR\,1716 (Sect.~\ref{2mass}). For a
quantitative derivation of fundamental parameters by means of a comparison with GCs, the isochrones 
were first set to the GCs according to the fundamental data taken from Harris (1996, and 
the 2003 update\footnote{http://physun.physics.mcmaster.ca/Globular.html}). Subsequently, we 
searched for the best overall match between the $\rm GC+isochrone$ and FSR\,1716 sequences. 
The results are shown in the middle panels of Fig.~\ref{fig8}. As a further test, we also
searched for solutions with the same isochrones and template GCs as above, but assuming a 
brighter MSTO for FSR\,1716, located at $\jj\approx13.5$. These tentative fits are sown in the
bottom panels of Fig.~\ref{fig8}\footnote{Although at approximately the same distance from the 
Sun and absorbed by a similar reddening value, the 2MASS cutoffs of M\,4 and NGC\,6397 are somewhat 
different. This occurs basically because M\,4 is projected against the bulge ($\ell\approx351\degr$), 
while NGC\,6397 lies at the outskirts ($\ell\approx338\degr$).}. Compared to NGC\,6397 (bottom-left
panel), the general fit is acceptable, although the RGB of FSR\,1716 is somewhat redder, which might 
suggest a higher metallicity. However, the fit with the more metal-rich GC M\,4 (bottom-right) is 
obviously poorer. Besides the red giants which are too red as compared to M\,4, the sub-giant 
branch morphologies become critically different. In the case of the M\,4-like solution, FSR\,1716 
would be as close ($\ds\approx0.6$\,kpc) as an old OC. In the case of the NGC\,6397 solution, 
FSR\,1716 would be the nearest ($\ds\approx0.5$\,kpc) GC so far discovered. 

The above comparisons show that a very low metallicity, like that of NGC\,6397, is not compatible 
with the giant branch of FSR\,1716, since it is steeper and bluer in NGC\,6397. M\,4, on the other
hand, provides a better match for the giant branch. What makes this possibility less likely is that 
the few extreme blue stars in the CMD of FSR\,1716 do not have counterpart in either NGC\,6397 or
M\,4, whose blue HBs are redder and well distributed in colour. The CMD shape for $13\la\jj\la15$ 
in FSR\,1716 is more compatible with the MSTO and MS of an old OC than the matched subgiant branches 
of the two GCs. However, the evidence of a Palomar-like (i.e. low-mass and loose structure) GC
(\citealt{Pap11GCs}), although marginal, cannot be ruled out with the existing data.

We call attention to the fact that it is not an easy task to establish the nature of
a low-mass GC or a low-mass old OC based on a few candidate HB or giant clump stars. For
instance, NGC\,188 (and other old OCs) presents only a few UV-bright stars, as shown by, 
e.g. \citet{Landsman98}. In fact, both low-mass GCs and low-mass old OCs can hardly have 
any HBs or clump giants, including NGC\,188 with only 5 clump giants (\citealt{Dinescu1995}; 
and \citealt{Landsman98}). As examples of low-mass GCs with a small number of UV-bright stars
we quote Palomar\,13 (\citealt{Cote2002}) with only a few, and AM-4 (\citealt{InCar87}) with 
none. 

Given the above circumstances, deeper observations would be necessary for a more conclusive 
definition on the nature of FSR\,1716 
as a GC (as originally suggested by \citealt{FSRcat}) or a very old OC, with a metallicity that 
can be sub Solar.

\subsection{CMD morphology of Cz\,23}
\label{NatureCz23}

The CMD features of Cz\,23 (bottom panels of Fig.~\ref{fig7}) indicate a younger age than FSR\,1716. 
Indeed, its field-decontaminated CMD morphology is similar to that of the old OC M\,67. The age of 
M\,67, as derived from near-infrared photometry, is $\sim3.2$\,Gyr (\citealt{M67}, and references 
therein). More recent determinations based on spectroscopy of individual stars put the age at $\sim4$\,Gyr 
(e.g. \citealt{Giampapa06}). In addition, the MSTO of Cz\,23 seems to be at a slightly fainter magnitude 
range, which indicates an age older than that of M\,67.

Thus, to probe the age and metallicity of Cz\,23, we show in Fig.~\ref{fig9} fits with the 3, 5, 
and 6\,Gyr isochrones with a fixed Solar metallicity (top panels), and the sub-Solar metallicity  
$[Fe/H]\sim-0.5$ (bottom). The fits are required to provide a similar description of the MS. It is
clear that, under this condition, the 3\,Gyr isochrone fails to reproduce the giant clump, in both 
metallicity ranges. The 5 and 6\,Gyr isochrones, on the other hand, provide acceptable MS and clump 
fits, which confirms the old age of Cz\,23. The 6\,Gyr solutions are obtained with an essentially 
zero reddening applied to the isochrones, which implies that older isochrones would require negative
correction. This argument constrains the age of Cz\,23 to the range $\approx4$ to $\approx6$\,Gyr.

\subsection{The adopted age/metallicity solutions}
\label{adoptFit}

\subsubsection{FSR\,1716}
\label{ageFSR}

Most of the arguments drawn above favour the old OC interpretation for the nature of
FSR\,1716. Since metal-poor OCs are preferentially distributed in the outer Galaxy 
(e.g. \citealt{Friel95}), and FSR\,1716 is located inside the Solar circle, its probable 
metallicity should be around the Solar value.

Based on the above, we take the 7\,Gyr, Solar metallicity isochrone as representative of
the stellar sequences of FSR\,1716, with an uncertainty of about $\pm1$\,Gyr. The 
corresponding {\em best-fit} was obtained with the reddening $\ejh=0.63\pm0.02$,
and observed distance modulus $\mMJ=11.1\pm0.2$. This solution is shown in the top panels of Fig.~\ref{fig7}.
Taking into account fit uncertainties we derive the age $7\pm1$\,Gyr, reddening $\ebv=2.0\pm0.1$ or,
equivalently (Sect.~\ref{2mass}) $A_V=6.3\pm0.2$, the absolute distance modulus $\mMo=9.36\pm0.21$,
and the distance from the Sun $\ds=0.8\pm0.1$\,kpc. Thus, for $\rs=7.2$\,kpc, the Galactocentric distance
of FSR\,1716 is $\dgc=6.6\pm0.2$\,kpc, which puts it $\approx0.6$\,kpc inside the Solar circle. 

Alternatively, we consider as well the less-likely possibility of a GC. In this case, the 
{\em best-fit} of the 12\,Gyr isochrone (based on the similarity with the M\,4 CMD morphology - 
Fig.~\ref{fig8}) is obtained with $\mMJ=13.5\pm0.3$, $\ejh=0.63\pm0.04$, $\ebv=2.0\pm0.1$ and
$A_V=6.3\pm0.2$, $\mMo=11.76\pm0.32$, $\ds=2.3\pm0.3$\,kpc, and $\dgc=5.4\pm0.2$\,kpc, thus 
$\approx1.8$\,kpc inside the Solar circle. 

\subsubsection{Cz\,23}
\label{ageCz}

With similar arguments as those used to estimate the age of FSR\,1716, we take the 5\,Gyr
($\pm1$\,Gyr) isochrone to represent the age of Cz\,23.  Since both metallicity ranges 
($[Fe/H]\sim-0.5$ and $0.0$) produce similar CMD fits (Fig.~\ref{fig9}), for simplicity, 
we also work with the Solar metallicity isochrone.

Thus, fundamental parameters of Cz\,23 are $\ejh=0.00\pm0.01$, $\ebv=0.00\pm0.03$ and 
$A_V=0.0\pm0.1$, $\mMJ=\mMo=12.0\pm0.1$, $\ds=2.5\pm0.2$\,kpc, and $\dgc=9.7\pm0.2$\,kpc; 
Cz\,23 lies $\approx2.5$\,kpc outside the Solar circle. This solution is shown in the bottom 
panels of Fig.~\ref{fig7}.

\section{Cluster structure}
\label{struc}

Structural parameters are derived by means of the RDPs, defined as the projected radial distribution
of the number density of stars around the cluster centre.

\begin{figure}
\resizebox{\hsize}{!}{\includegraphics{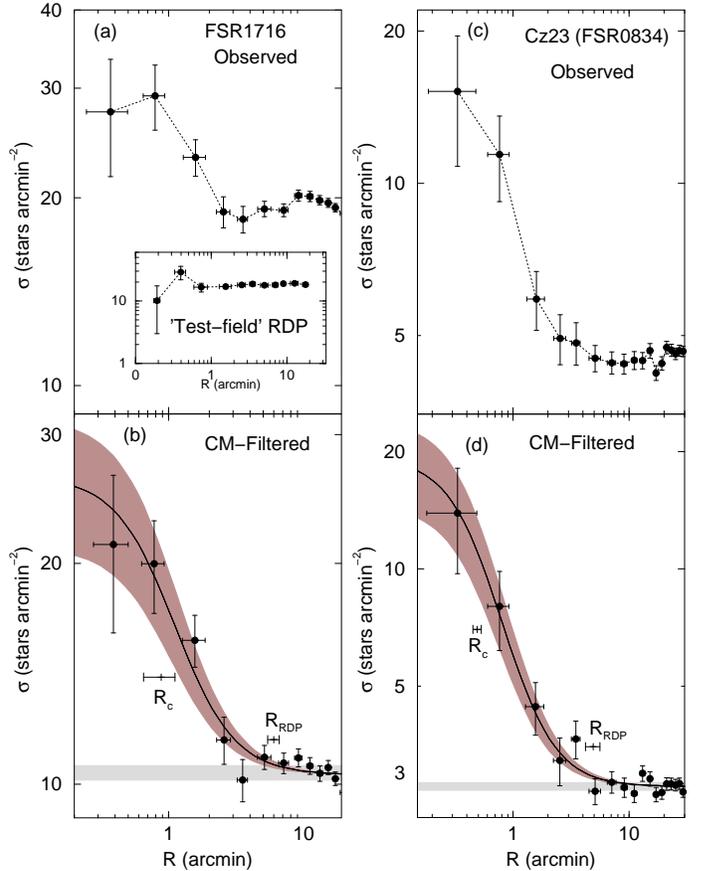}}
\caption[]{Stellar RDPs (filled circles) of FSR\,1716 (left panels) and Cz\,23 (right). RDPs built with
the observed (raw) photometry are shown in the top panels, while those built with colour-magnitude
filtered photometry are in the bottom panels. Solid line: best-fit King-like profile. Horizontal shaded
polygon: offset field stellar background level. Shaded regions: $1\sigma$ King fit uncertainty. The core
and cluster radii are indicated. Inset of panel (a): RDP of the test-field of FSR\,1716. Angular scale
is used.}
\label{fig10}
\end{figure}

Star clusters usually have RDPs that follow a well-defined analytical profile.
The most often used are the single mass, modified isothermal sphere (\citealt{King66}), the modified
isothermal sphere (\citealt{Wilson75}), and the power law with a core (\citealt{EFF87}). Each function
is characterised by different parameters that are related to cluster structure. However,
because the error bars in the present RDPs are significant (Fig.~\ref{fig10}), we use the
analytical profile $\sigma(R)=\sigma_{bg}+\sigma_0/(1+(R/R_c)^2)$, where $\sigma_{bg}$ is
the residual background density, $\sigma_0$ is the central density of stars, and \rc\ is the core
radius. This function is similar to that by \cite{King1962} to describe the surface
brightness profiles in the central parts of globular clusters.

\begin{table*}
\caption[]{Cluster structural parameters}
\label{tab3}
\renewcommand{\tabcolsep}{2.25mm}
\renewcommand{\arraystretch}{1.25}
\begin{tabular}{lcccccccccc}
\hline\hline
Cluster&$1\arcmin$&&$\sigma_{bg}$&$\sigma_0$&$\delta_c$&\rc&\rl&\rc&\rl\\
       &(pc)&&$\rm(stars\,arcmin^{-2})$&$\rm(stars\,arcmin^{-2})$&&(\arcmin)&(\arcmin)&(pc)&(pc)\\
(1)&(2)&&(3)&(4)&(5)&(6)&(7)&(8)&(9)\\
\hline
FSR\,1716$^\dagger$&0.216&&$10.5\pm1.0$&$16.2\pm4.8$&$2.6\pm0.5$&$0.88\pm0.23$&$6.0\pm0.6$&
   $0.19\pm0.05$&$1.3\pm0.2$&\\

FSR\,1716$^\ddagger$&0.653&&$10.5\pm1.0$&$16.2\pm4.8$&$2.6\pm0.5$&$0.88\pm0.23$&$6.0\pm0.6$&
   $0.57\pm0.15$&$3.9\pm0.4$&\\

Cz\,23   &0.728&&$2.8\pm0.1$&$16.5\pm4.7$&$7.0\pm1.7$&$0.49\pm0.04$&$4.9\pm0.7$&$0.36\pm0.08$ 
   &$3.6\pm0.4$\\
\hline
\end{tabular}
\begin{list}{Table Notes.}
\item Col.~2: arcmin to parsec scale. King profile is expressed as
$\sigma(R)=\sigma_{bg}+\sigma_0/(1+(R/R_c)^2)$. To minimise degrees of freedom in RDP fits, $\sigma_{bg}$ 
was kept fixed (measured in the respective comparison fields) while $\sigma_0$ and \rc\ were allowed to 
vary. Col.~5: cluster/background density contrast ($\delta_c=1+\sigma_0/\sigma_{bg}$), measured in CM-filtered
RDPs. Cols.~6-9, core and cluster radii in angular and absolute units. Parameters of FSR\,1716 are derived 
for the OC ($\dagger$) or GC ($\ddagger$) interpretation (Sect.~\ref{age}).
\end{list}
\end{table*}

The RDPs are built with colour-magnitude filtered photometry to minimise noise. The most probable cluster
sequences are isolated by means of colour-magnitude filters, which are used to exclude stars with colours
different from those of the assumed cluster sequence. They are wide enough to include cluster MS and
evolved star colour distributions, as well as the $1\sigma$ photometric uncertainties. Colour-magnitude
filter widths should also account for formation or dynamical evolution-related effects, such as enhanced
fractions of binaries (and other multiple systems) towards the central parts of clusters, since such systems
tend to widen the MS (e.g. \citealt{HT98}; \citealt{Kerber02}; \citealt{BB07}; \citealt{N188}). The 
colour-magnitude filters
for the present OCs are shown in the left panels of Fig.~\ref{fig7}. Residual field stars with
colours similar to those of the cluster are expected to remain inside the colour-magnitude filter region.
They affect the intrinsic stellar RDP in a way that depends on the relative densities of field and
cluster stars. The contribution of the residual contamination to the observed RDP is statistically 
subtracted by means of the field. In practical terms, the use of colour-magnitude
filters in cluster sequences enhances the contrast of the RDP with respect to the background,
especially for clusters in dense fields (e.g. \citealt{BB07}).

Oversampling near the centre and undersampling at large radii are avoided by using rings
of increasing width with distance from the centre. A typical set of ring widths is
$\Delta\,R=0.5,\ 1,\ 2,\ {\rm and}\ 5\arcmin$, respectively for $0\arcmin\le R<1\arcmin$,
$1\arcmin\le R<4\arcmin$, $4\arcmin\le R<10\arcmin$, and $10\arcmin\le R<30\arcmin$. The number and
width of the rings can be set to produce RDPs with adequate spatial resolution and as low as
possible $1\sigma$ Poisson errors. The residual background level of each RDP corresponds to the average
number of colour-magnitude filtered stars measured in the field. The $R$ coordinate (and
 uncertainty) of each ring corresponds to the average position and standard deviation of the
stars inside the ring.

The RDPs of FSR\,1716 and Cz\,23 are given in Fig.~\ref{fig10}. Because of the uncertainties
associated to the age (and metallicity) derivation (Sect.~\ref{age}), which propagate to the absolute 
value of the 
structural parameters, RDPs in Fig.~\ref{fig10} are shown in angular scale. Besides the RDPs built 
with the colour-magnitude filters,
we show, for illustrative purposes, those produced with the observed (raw) photometry. Especially
for FSR\,1716, minimisation of the number of non-cluster stars by the colour-magnitude filter resulted in
a RDP with a higher contrast with respect to the background. Fits of the King-like profile
were performed with a non-linear least-squares fit routine that uses errors as weights. To minimise degrees
of freedom, $\sigma_0$ and \rc\ were derived from the RDP fit, while $\sigma_{bg}$ is measured in the
field. These values are given in Table~\ref{tab3}, and the best-fit solutions are
superimposed on the colour-magnitude filtered RDPs (Fig.~\ref{fig10}). Table~\ref{tab3}
also provides structural parameters in absolute units, computed with the cluster distances derived in
Sect.~\ref{adoptFit}. Because of the 2MASS photometric
limit, which in most cases corresponds to a cutoff for stars brighter than $\jj\approx16$, $\sigma_0$
should be taken as a lower limit to the actual central number density. The adopted King-like function
describes well the RDPs along the full radius range, within uncertainties.

To complete the structural description of the clusters we also estimate the cluster radius (\rl) and
uncertainty by visually comparing the RDP level (and fluctuations) with the background. The cluster
radius corresponds to the distance from the cluster centre where RDP and background are statistically
indistinguishable (e.g. \citealt{DetAnalOCs}, and references therein). For practical purposes, most 
of the cluster stars are contained within $\rl$. Note that \rl\ should not be mistaken for the tidal 
radius. Tidal radii are derived from King fits to RDPs, which depend on wide fields and adequate
Poisson errors. For instance, in populous and relatively high Galactic latitude OCs
such as M\,67, NGC\,188, and NGC\,2477, cluster radii are a factor $\sim0.5 - 0.7$ of the respective
tidal radii (\citealt{DetAnalOCs}). The cluster radii of the present objects are given in cols.~6-7
(angular scale) and 8-9 (absolute scale) of Table~\ref{tab3}. If cluster and tidal radii of FSR\,1716
and Cz\,23 are similarly related as in the bright OCs, the lower-limit of the radial range
taken as comparison field (Sect.~\ref{Decont_CMDs}) is located beyond the probable tidal radius. This, in
turn, minimises the probability of cluster members at large radii, e.g. in the cluster halo, to be
considered as field stars by the decontamination algorithm.

Compared to the distribution of core radius derived for a sample of relatively nearby OCs by 
\citet{Piskunov07}, FSR\,1716 (especially the OC interpretation) and Cz\,23 occupy the small-\rc\
tail. Assuming that the tidal radius is $\sim2\times\rl$, Cz\,23 and FSR\,1716 (GC interpretation)
would be around the median value of the distribution, while the OC interpretation of FSR\,1716 again
would be at the small-tidal radius tail. 

Table~\ref{tab3} (col.~5) provides the density contrast parameter $\delta_c=1+\sigma_0/\sigma_{bg}$.
Since $\delta_c$ is measured in colour-magnitude-filtered RDPs, it may not correspond to the
visual contrast produced by observed stellar distributions in images. FSR\,1716 presents a
low contrast in the 2MASS \ks\ image (Fig.~\ref{fig1}) but, because most of the non-cluster stars have
been excluded by the colour-magnitude filter, the corresponding RDP has a relatively high density contrast, 
$\delta_c\approx2.6$. Obviously, the high-contrast OC Cz\,23 (Figs.~\ref{fig1} and
\ref{fig10}) is reflected
in the high value of the density-contrast parameter $\delta_c\approx7.0$. Interestingly, FSR\,1716 and
Cz\,23 are projected against almost opposite directions (Table~\ref{tab1}). Accordingly, the background
(including foreground) stellar contribution turns out to be $\rm\approx11\,stars\,arcmin^{-2}$ in the
direction of FSR\,1716, and $\rm\approx2.8\,stars\,arcmin^{-2}$ towards Cz\,23.

Finally, the RDP of the `test-field' of FSR\,1716 is shown in the inset of panel (a) of Fig.~\ref{fig10}.
As expected from the CMD analysis (Sect.~\ref{EastField}), the RDP is uniform from the centre to the
borders of the field, characterised essentially by Poisson fluctuations. The featureless `test-field'
RDP is low and flat, as compared the cluster RDP of FSR\,1716 (panels a and c).

\section{Mass function and cluster mass estimate}
\label{MF}

Cluster mass functions $\left(\phi(m)=\frac{dN}{dm}\right)$ are built following the methods presented
in \citet{DetAnalOCs} (and references therein). We build them with colour-magnitude filtered photometry, the
3 2MASS bands separately, and the mass-luminosity relations obtained from the Padova isochrones
and distances from the Sun adopted for FSR\,1716  (Sect.~\ref{ageFSR}) and Cz\,23 (Sect.~\ref{ageCz}).
Further details on MF construction are given in \citet{FaintOCs}.
The effective magnitude range over which MFs are computed is that where clusters present an excess of stars
over the field. In both cases it begins below the MSTO and ends at a faint magnitude
limit brighter than that stipulated by the 2MASS completeness limit (Sect.~\ref{2mass}). The effective MS
stellar mass ranges are $0.83\leq m(\ms)\leq1.1$ (FSR\,1716), and $0.94\leq m(\ms)\leq1.2$ (Cz\,23).

\begin{figure}
\resizebox{\hsize}{!}{\includegraphics{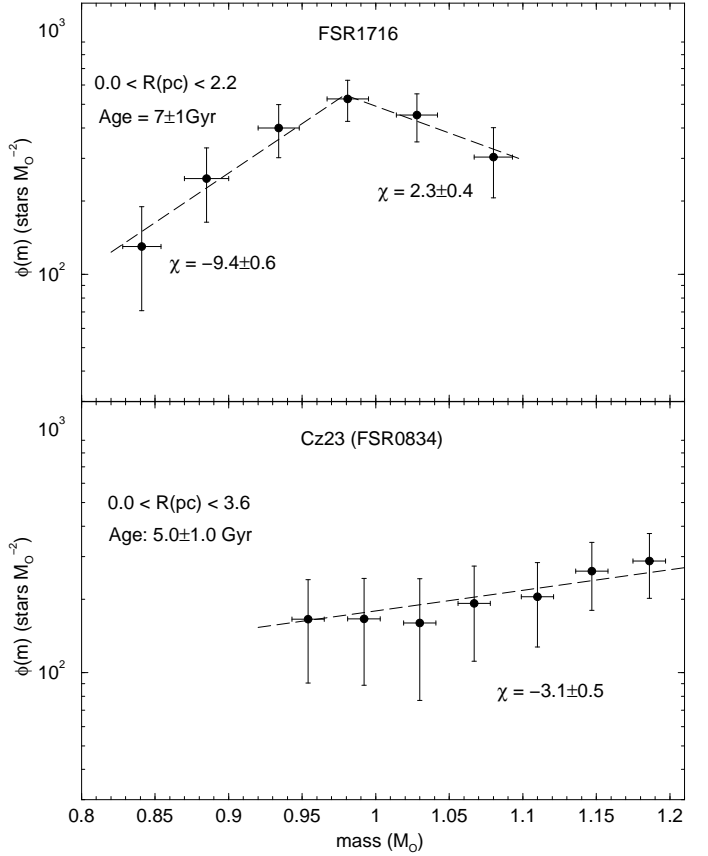}}
\caption[]{Mass function computed for the whole region of FSR\,1716 (top panel) and Cz\,23 (bottom).
The MF of FSR\,1716 can be represented by the broken power-law $\phi(m)\propto m^{-(1+\chi)}$, with
$\chi=-9.4\pm0.6$ for $m\le0.98\,\ms$, and $\chi=2.3\pm0.4$ for $0.98<m(\ms)<1.1$. The MF drop begins 
at $m\approx0.98\,\ms$, which corresponds to $\jj\approx14.5$. The MF of Cz\,23 can be described by 
a power-law with slope $\chi=-3.1\pm0.5$.}
\label{fig11}
\end{figure}

The MFs computed for the whole cluster region ($R<\rl$) are shown in Fig.~\ref{fig11}. Both cases suggest
important depletion of the low-mass content, suggesting advanced dynamical evolution, especially
in the case of FSR\,1716. The drop in the MF of FSR\,1716 begins at $m\la0.98\,\ms$, which corresponds to 
$\jj\approx14.5$, more than 1 mag brighter than the 2MASS 99.9\% completeness limit (Sect.~\ref{2mass}) at
the cluster position. This suggests that the drop may be real, although we cannot rule out crowding to account for 
part of the important MF drop, since the 2MASS completeness limits refer to the field.
Bearing in mind this caveat, we fit the MF of FSR\,1716 with a two-segment function, $\phi(m)\propto m^{-(1+\chi)}$, 
with $\chi=-9.4\pm0.6$ in the mass range $m\le0.98\,\ms$, and $\chi=2.3\pm0.4$ for $0.98<m(\ms)<1.1$. 
The high-mass range slope is, within the uncertainty, somewhat steeper than the $\chi=1.35$ of 
\citet{Salpeter55} Initial Mass Function (IMF). However, the low-mass range slope is much flatter than
Salpeter's IMF. The drop in the MF of FSR\,1716 (at $m\approx1\,\ms$) agrees with one of the breaks present 
in the universal IMF of \citet{Kroupa2001}, which assumes increasing flattening towards low-mass stars.
This IMF is described by the slopes $\chi=0.3\pm0.5$ for the range $0.08\leq m(\ms)\leq0.5$ and 
$\chi=1.3\pm0.3$ for $0.5\leq m(\ms)\leq1.0$. The low-mass range of the MF of FSR\,1716 is flatter
than Kroupa's IMF, which again points to advanced dynamical evolution, crowding or, more probably,
a combination of both.

Although flat, the MF of Cz\,23 is more monotonic than that of FSR\,1716, and can be described by a
single power-law, with the slope $\chi=-3.1\pm0.5$, also flatter than Salpeter's or Kroupa's IMF.

The number of observed MS and evolved stars in FSR\,1716 is derived by counting the number
of stars (in the background-subtracted colour-magnitude filtered photometry) that show up in the
corresponding magnitude ranges, $13.4<\jj<15.4$ for the MS and $\jj<13.4$ for the evolved stars.
There are $n_{MS}=99\pm5$ and $n_{evol}=53\pm7$, MS and evolved stars, respectively; the corresponding
mass values are $m_{MS}=97\pm5\,\ms$ and $m_{evol}=58\pm8\,\ms$. The evolved star mass corresponds to
$n_{evol}$ multiplied by the stellar mass at the MSTO, $m_{TO}=1.1\,\ms$. To estimate the total stellar
mass we extrapolate the observed MF down to the H-burning mass limit ($0.08\,\ms$). Because of the very
flat slope at the low-mass range, the extrapolated values are similar to the observed ones,
$n_{extr}=160\pm8$ and $m_{extr}=160\pm9\,\ms$.

The number of MS and evolved stars in Cz\,23 is $n_{MS}=54\pm4$ and $n_{evol}=5\pm3$, and the
respective masses are $m_{MS}=60\pm4\,\ms$ and $m_{evol}=7\pm4\,\ms$; the extrapolated values are
$n_{extr}=120\pm8$ and $m_{extr}=115\pm7\,\ms$.

Presently, both FSR\,1716 and Cz\,23 appear to be low-mass OCs, even when stars less massive than the
observed range are taken into account. Most of the original mass must have been lost to
the field because of dynamical effects along the several Gyrs since their formation (Sect.~\ref{Discus}).

\section{Discussion}
\label{Discus}

The range of acceptable isochrone solutions (Sect.~\ref{age}), together with the statistical tests 
(Sect.~\ref{Decont_CMDs}) applied to FSR\,1716 and Cz\,23, i.e. {\em (i)} the decontamination algorithm, 
{\em (ii)} the integrated and per magnitude \ns\ parameter, and {\em (iii)} the ratio of \nc\ to \sFS, 
produce results consistent with both objects being old OCs. Examination of photometric and structural 
properties of the offset 'test-fields' for FSR\,1716 and Cz\,23, also leads to the same conclusion.

With the analyses of the preceding sections, we provide fundamental and structural parameters, 
most of which for the first time for FSR\,1716 and Cz\,23. We now use these parameters to put both 
clusters into perspective, by comparing some of their properties with those of a set of well-studied 
OCs. 

We compare the structural parameters, computed with the adopted isochrone solutions for FSR\,1716
(Sect.~\ref{ageFSR}) and Cz\,23 (Sect.~\ref{ageCz}), with those of a reference sample of nearby OCs 
with ages in the range $70-7\,000$\,Myr and masses within $400-5\,300$\,\ms\ (\citealt{DetAnalOCs}). 
To the original reference sample were added the young OCs NGC\,6611 (\citealt{N6611}) and NGC\,4755 
(\citealt{N4755}). Clusters are distinguished according to total mass (lower or higher than 1\,000\,\ms).
\citet{DetAnalOCs} discuss parameter correlations in the reference sample. For completeness,
both sets of parameters for FSR\,1716 (Table~\ref{tab3}), corresponding to the old OC and GC possibilities, 
are considered separately.

\begin{figure}
\resizebox{\hsize}{!}{\includegraphics{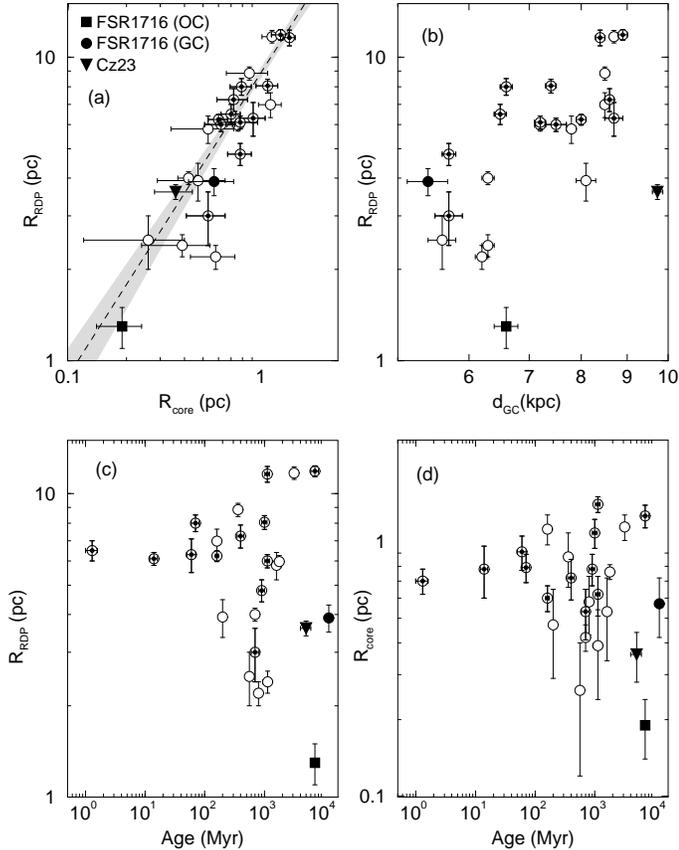}}
\caption[]{Relations involving OC structural and fundamental parameters. Circles: nearby OCs. 
Dotted circles: massive OCs ($>1\,000\,\ms$). Two sets of parameters are shown for FSR\,1716, 
according to the old OC and GC solutions.}
\label{fig12}
\end{figure}

Core and cluster radii of the reference sample relate as $\rl=(8.9\pm0.3)\times R_c^{(1.0\pm0.1)}$
(panel a), which suggests a similar scaling for both kinds of radii, at least for the sampled
ranges of $0.25\la\rc(pc)\la1.5$ and $2\la\rl(pc)\la15$. Within uncertainties, FSR\,1716 (both 
interpretations) and Cz\,23 fit tightly into that relation. 

The reference OCs appear to follow the trend of increasing cluster radii with Galactocentric 
distance (panel b), a dependence previously suggested by, e.g. \citet{Lynga82}. To explain the 
increase of GC radii with Galactocentric distance, \citet{vdB91} suggested that part of the 
relation may be primordial, in the sense that the higher molecular gas density in central 
Galactic regions may have produced clusters with small radii. After formation, mass loss 
associated to stellar and dynamical evolution (such as mass segregation and evaporation), 
together with tidal interactions with the Galactic potential and giant molecular clouds, also 
contribute to the depletion of star clusters, especially the low-mass and centrally located 
ones (Sect.~\ref{Intro}). The cluster radius of FSR\,1716 (GC interpretation) is consistent with 
the assumed relation. On the other hand, Cz\,23, and the OC interpretation of FSR\,1716, appear 
exceedingly small for their Galactocentric distance, which is consistent with the several Gyrs
of depletion. A similar dependence with Galactocentric distance holds as well for \rc, because of 
the relation with \rl implied by panel (a).

In panels (c) and (d) of Fig.~\ref{fig12} we compare the cluster and core radii of FSR\,1716 and 
Cz\,23 with those of the reference sample in terms of age. The locus occupied by Cz\,23 in both 
panels is consistent with the corresponding radii measured in low-mass, old OCs. A similar 
conclusion applies to the GC interpretation of FSR\,1716 (which again would be consistent with the 
structural parameters of Palomar-like GCs - \citealt{Pap11GCs}). The significantly small radii 
of the OC interpretation of FSR\,1716, especially \rl, are consistent with those of OCs that have 
suffered severe depletion effects for long periods (e.g. \citealt{DetAnalOCs}; \citealt{OldOCs},
and references therein).   

\begin{figure}
\resizebox{\hsize}{!}{\includegraphics{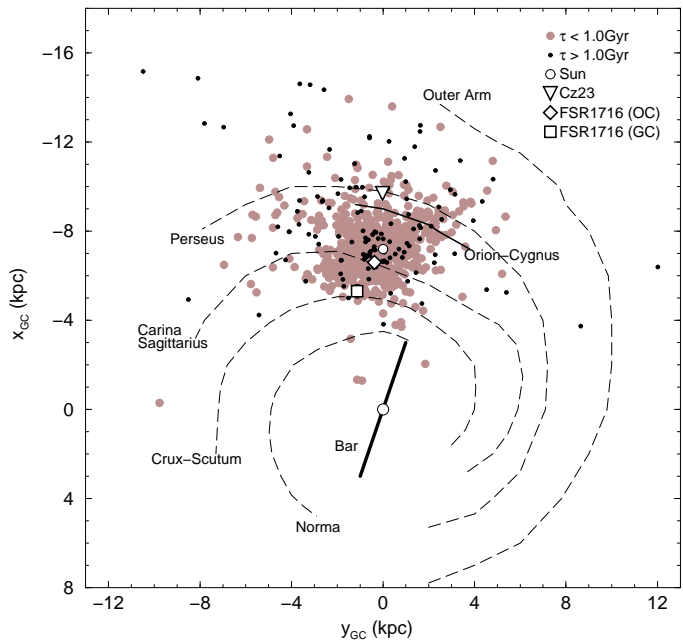}}
\caption[]{Galactic position of FSR\,1716 and Cz\,23 compared to the WEBDA OCs with ages 
younger (shaded circles) and older (filled circles) than 1\,Gyr. The schematic projection 
of the Galaxy is seen from the North pole, with 7.2\,kpc as the Sun’s distance to the 
Galactic centre. Main structures are identified.}
\label{fig13}
\end{figure}

Another evidence in favour of dynamical effects affecting cluster size comes from the positions 
of FSR\,1716 and Cz\,23 in the Galactic plane (Fig.~\ref{fig13}). Milky Way's spiral arm structure 
is based on \citet{GalStr} and \citet{DrimSper01}, derived from HII regions and molecular clouds 
(e.g. \citealt{Russeil03}). The Galactic bar is shown with a 14\degr\ orientation and 3\,kpc in 
length (\citealt{Freuden98}; \citealt{Vallee05}). For comparison purposes, we also include WEBDA 
clusters younger and older than 1\,Gyr. As discussed in Sect.~\ref{Intro}, old OCs distribute 
preferentially outside the Solar circle.

Interestingly, both clusters occur close to spiral arms, Cz\,23 on Perseus, and FSR\,1716 on
Carina-Sagittarius (OC interpretation) or Crux-Scutum (GC). Since they are located close to
the plane (Table~\ref{tab1}), they may have interacted with the arms, especially by means of 
encounters with giant molecular clouds\footnote{Molecular clouds more massive than $\sim10^6\,\ms$ 
are found in the Solar neighbourhood (e.g. \citealt{Solo87}).}. As discussed above, part of the
small sizes of FSR\,1716 and Cz\,23 can be accounted for by collisions with such structures 
(e.g. \citealt{Wielen71}; \citealt{Wielen91}; \citealt{Gieles06}; \citealt{GAP07}). A similar
effect was recently observed to occur with some small OCs located close to the Local and 
Orion-Cygnus Arms (\citealt{AntiCent}).

Finally, both objects have MFs that appear to be very eroded and flat which, given the discussion on
structural parameter above, reflects the consequences of several Gyrs of the relentless dynamical 
effects and mass loss due to stellar evolution. Besides, both objects are low-mass clusters, with 
less than $\sim200\,\ms$ presently stored in stars. They must have been formed as more massive open 
clusters to have survived for so long in the Galaxy. 

\section{Concluding remarks}
\label{Conclu}

In this paper we study the nature of two stellar overdensities included in the catalogue of candidate 
star clusters 
of \citet{FSRcat}, FSR\,1716 and FSR\,834. The former was suggested to be a globular cluster candidate 
by \citet{FSRcat}, while the latter has the OC Cz\,23 as optical counterpart. The analyses are based 
on field-star decontaminated 2MASS CMDs and stellar radial density profiles, with algorithms previously 
constructed by our group. Fundamental and structural parameters of the clusters are derived.

We present consistent evidence (e.g. CMD morphology, statistical tests, structural parameters,
mass-function slope, and comparison with nearby OCs) that both objects are old star clusters.

FSR\,1716 is significantly absorbed ($\aV\approx6.3$) and projected not far from the bulge. Its 
field-decontaminated CMD morphology is very similar to that of the $\sim7$\,Gyr well-known OC NGC\,188. 
Indeed, its CMD can be well represented by isochrones with ages older than $\sim6$\,Gyr, both of Solar 
and sub-Solar ($[Fe/H]\sim-0.5$) metallicity. We adopted the former as the metallicity of FSR\,1716, 
because of its relatively central location. Alternatively, we cannot rule out the possibility that 
FSR\,1716 is a low-mass, loose (Palomar-like) globular cluster. FSR\,1716 is located inside the Solar 
circle, $\approx0.6$\,kpc (in the case of an OC) or $\approx1.8$\,kpc (GC). 

The CMD morphology of Cz\,23 is consistent with that of the $\sim4$\,Gyr old OC M\,67. Similarly to 
FSR\,1716, it can be well represented by Solar and sub-Solar metallicity isochrones, but with ages 
in the range $4-6$\,Gyr. With the $\approx5$\,Gyr and Solar metallicity solution, we find that Cz\,23 
is projected nearly towards the anti-centre, located $\approx2.5$\,kpc outside the Solar circle. 

The core and cluster radii of FSR\,1716 and Cz\,23 are small when compared to a set of open clusters 
in the Solar neighbourhood. However, such radii are comparable with those of other OCs of similar old 
age. Besides, the mass functions appear to be much flatter than Salpeter's IMF, especially FSR\,1716, 
which seems to present an increasing depletion in the number of low-mass stars. As a consequence, the 
total masses presently stored in stars in both clusters are lower than $\sim200\,\ms$. Such low values 
probably reflect the several Gyr-long period of mass loss due to stellar evolution, tidal interactions 
with the bulge (possibly in the case of FSR\,1716), disk and giant molecular clouds. Actually, because 
of its low mass content and flat mass function, Cz\,23 may be evolving into an open cluster remnant 
(e.g. \citealt{PB07}).

Comprehensive catalogues of star cluster candidates, such as that of \citet{FSRcat}, should be
further explored with field-star decontamination algorithms and other tools, so that the nature 
of the candidates can be probed and the age derived. It is remarkable how the decontamination 
tool unveiled the intrinsic CMD sequences of FSR\,1716, separating it from the crowded 
field population. Consequently, the characterisation of FSR\,1716 and Cz\,23 as OCs older than 
$\sim4$\,Gyr represents an important increase ($\approx10\%$) to the known population of such 
objects in the Galaxy. In particular, FSR\,1716 is the most recent addition to the 8 open clusters 
older than 6\,Gyr so far identified (WEBDA). In this sense, Cz\,23 (FSR\,834), and especially 
FSR\,1716, can be considered as rare examples of extreme dynamical survivors in disk-regions where 
most open clusters are short-lived. 

\section*{Acknowledgements}
We thank the anonymous referee for helpful suggestions.
This publication makes use of data products from the Two Micron All Sky Survey, which
is a joint project of the University of Massachusetts and the Infrared Processing and
Analysis Centre/California Institute of Technology, funded by the National Aeronautics
and Space Administration and the National Science Foundation. This research has made 
use of the WEBDA database, operated at the Institute for Astronomy of the University
of Vienna. We acknowledge support from the Brazilian Institution CNPq.

\label{lastpage}
\end{document}